\documentclass[twocolumn]{aastex631}

\let\tablenum\relax
\usepackage{siunitx}
\usepackage{physics}
\usepackage{comment}
\usepackage{tabularx}
\usepackage{color}

\begin{document}

\title{Phase-Dependent Spectral Shape Changes in the Ultraluminous X-Ray Pulsar NGC 5907 ULX1}

\author{Daiki Miura}
\affiliation{Department of Physics, Graduate School of Science, The University of Tokyo, 7-3-1 Hongo, Bunkyo-ku, Tokyo 113-0033, Japan}
\affiliation{Institute of Space and Astronautical Science (ISAS), Japan Aerospace Exploration Agency (JAXA), 3-1-1 Yoshinodai, Chuo-ku, Sagamihara, Kanagawa 252-5210, Japan}

\author[0000-0001-7773-9266]{Shogo B. Kobayashi}
\affiliation{Department of Physics, Tokyo University of Science, 1-3 Kagurazaka, Shinjuku-ku, Tokyo 162-8601, Japan}

\author[0000-0002-5092-6085]{Hiroya Yamaguchi}
\affiliation{Institute of Space and Astronautical Science (ISAS), Japan Aerospace Exploration Agency (JAXA), 3-1-1 Yoshinodai, Chuo-ku, Sagamihara, Kanagawa 252-5210, Japan}
\affiliation{Department of Physics, Graduate School of Science, The University of Tokyo, 7-3-1 Hongo, Bunkyo-ku, Tokyo 113-0033, Japan}

\correspondingauthor{Daiki Miura}
\email{dikmur0611@g.ecc.u-tokyo.ac.jp}



\begin{abstract}
Discovery of coherent pulsations from several ultraluminous X-ray pulsars (ULXPs) has provided direct evidence of super-critical accretion flow. However, geometrical structure of such accretion flow onto the central neutron star remains poorly understood. 
NGC 5907 ULX1 is one of the most luminous ULXPs with the luminosity exceeding $10^{41}$\,erg\,s$^{-1}$. 
Here we present a broadband X-ray study of this ULXP using 
the data from simultaneous observations with {\rm XMM-Newton} and {\rm NuSTAR} conducted in July 2014.
The phase-resolved spectra are well reproduced by a model consisting of 
a multicolor disk blackbody emission with a temperature gradient of $p=0.5$ ($T\propto r^{-p}$) and a power law with an exponential cutoff.
The disk component is phase-invariant, and has an innermost temperature of $\sim$\,\SI{0.3}{keV}. Its normalization suggests a relatively low inclination angle of the disk, in contrast to the previous claim in other literature.
The power law component, attributed to the emission from the accretion flow inside the magnetosphere of the neutron star, indicates phase-dependent spectral shape changes; the spectrum is slightly harder in the pre-peak phase than in the post-peak phase.
This implies that the magnetosphere has an asymmetric geometry around the magnetic axis, and that hotter regions close to the magnetic pole become visible before the pulse peak.
\end{abstract}



\section{Introduction} \label{sec:intro}
Ultraluminous X-ray sources (ULXs: \citealt{makishima_nature_2000}) are point-like sources with their apparent luminosity of $L_X > 10^{39}$\,erg\,s$^{-1}$, often found in the off-nuclear regions of nearby galaxies (e.g., \citealt{king_ultraluminous_2023,pinto_ultra-luminous_2023} for recent reviews). 
Since the luminosity exceeds the Eddington limit for a typical stellar-mass black hole, ULXs have once been considered to be candidates for intermediate mass black holes (e.g., \citealt{colbert_nature_1999}; \citealt{zampieri_low-metallicity_2009}).
To date, more than 1800 ULXs have been discovered \citep{walton_multimission_2021}. 

The situation has dramatically changed by the discovery of coherent pulsations from the ULX M82 X-2, indicating that the system consists of a neutron star accreting at the super-Eddington rate \citep{bachetti_ultraluminous_2014}. 
Following this discovery, similar pulsations have been detected from other ULXs, 
e.g., NGC 7793 P13 \citep{furst_discovery_2016,israel_discovery_2017}, NGC 5907 ULX1 \citep{israel_accreting_2017}, NGC 300 ULX-1 \citep{carpano_discovery_2018}, NGC 1313 X-2 \citep{sathyaprakash_discovery_2019}. There are approximately 10 ultraluminous X-ray pulsars (ULXPs) that have been identified to date.
Since the mass of a neutron star is at most $\sim$\,3$M_\odot$, the extremely high luminosity observed in the ULXPs provides direct evidence for the accretion rate exceeding the Eddington limit, offering an ideal site to study the poorly-understood nature of super-critical accretion flow. 

NGC 5907 ULX1 is one of the most luminous ULXPs, located on the disk of the edge-on spiral galaxy NGC 5907, the distance to which is $\SI{17.1}{Mpc}$ \citep{tully_cosmicflows-3_2016}. 
The source was first reported in the ULX catalog presented by \cite{walton_2xmm_2011}. 
Observing this ULX with {\rm XMM-Newton} and {\rm NuSTAR}, 
\cite{israel_accreting_2017} discovered a coherent pulsations with the peak luminosity of $\sim \SI{e41}{erg.s^{-1}}$. 
The spin period and its first derivative obtained from the July 2014 observations were $P \sim \SI{1.137}{s}$ and $\dot{P}\sim \SI{-5e-9}{s.s^{-1}}$, respectively. 
They also revealed an energy-dependent phase shift in its pulse profiles, 
which indicates the presence of spectral variations as a function of the pulse phase. 
However, no quantitative report on the phase-dependent spectral parameters was given in their work.

The spectra of NGC 5907 ULX1 have been revisited in some other literature. 
\cite{walton_evidence_2018} performed phase-resolved spectroscopy, but the spectral shape of the pulsation component was assumed to be invariable with respect to the pulse phase. In other words, the energy-dependent phase shift reported by \cite{israel_accreting_2017} was not taken into account in their work. 
\cite{furst_spectral_2017} analyzed the phase-averaged spectra of this ULXP to investigate the properties of the accretion disk. From the relationship between the modeled bolometric luminosity and the observed flux of the disk component, they estimated the inclination angle of the disk to be $\sim$\,89$^\circ$, implying a nearly edge-on view of the system. This result is puzzling, if true, because in a super-critical accretion system, the accretion disk is thought to become geometrically thick to block the line of sight to the pulsating accretion flow inside the magnetosphere when viewed from the side. It should be noted, however, that in spectral analysis of typical ULXP, the spectral parameters of the disk and pulsation components are easily coupled with each other. Therefore, careful spectral modeling is essential to constrain the physical properties of both components. 

Motivated by the previous work, we revisit the X-ray spectra of NGC 5907 ULX1 obtained from the July 2014 observations simultaneously conducted by {\rm XMM-Newton} and {\rm NuSTAR}. 
We perform phase-resolved broadband spectroscopy by taking into account the phase-dependent spectral variations \citep{israel_accreting_2017} 
to gain insights into the nature of the accretion flow. 

The rest of this paper is structured as follows. Details of observations and data reduction are described in Section \ref{sec:obs}. We present analysis and results in Section \ref{sec:analysis} and discuss their implications in Section \ref{sec:discussion}. Finally, we conclude this study in Section \ref{sec:conclusion}.
The errors quoted in the text and table and error bars given in figures represent a 1$\sigma$ confidence level unless otherwise stated.

\section{observations and data reduction} \label{sec:obs}

In this work, we analyze the archival data of NGC 5907 ULX1 from the European Photon Imaging Camera (EPIC) aboard {\rm XMM-Newton} \citep{jansen_xmm-newton_2001} and the Focal Plane Module (FPM) aboard {\rm NuSTAR} \citep{harrison_nuclear_2013}. Both instruments have a large effective area and adequate temporal resolution, allowing us to perform high-quality phase-resolved spectroscopy in the wide X-ray band. 

\subsection{{\rm XMM-Newton}}
NGC 5907 ULX1 was observed by {\rm XMM-Newton} on 2014 July 9 with an effective exposure of $\sim$\,$\SI{38}{ks}$ for the EPIC-pn \citep{struder_european_2001} and EPIC-MOS \citep{turner_european_2001} detectors (ObsID: 0729561301).
Both instruments were operating in the Full Frame mode. 
Since the timing resolution of the EPIC-pn is much better than that of the EPIC-MOS ($\SI{73.4}{ms}$ and $\sim \SI{1}{s}$, respectively), we use only the former in the following analysis. 
We reprocessed the data using the \texttt{epchain} task in the {\rm XMM-Newton} Science Analysis System (SAS) v20.0.0. 
The cleaned event data were corrected to the barycenter of the solar system using the SAS tool \texttt{barycen} by assuming the solar system ephemeris DE200 \citep{standish_observational_1990}. We use only single- and double-pixel events. 
No significant flare was detected during the observation. 
Following \cite{israel_accreting_2017}, we adopted a source coordinate of $\text{RA}=15^\mathrm{h}15^\mathrm{m}58^\mathrm{s}.62$ and $\text{DEC}=+\ang{56;18;10}.3$, and extracted the source data from a $\ang{;;30}$-radius circular region around the source position. The background data were extracted from an annulus region with inner- and outer-radii of  $\ang{;;30}$ and $\ang{;;90}$, respectively.

\subsection{{\rm NuSTAR}}
{\rm NuSTAR} observed the source with an exposure of $\sim$\,$\SI{57}{ks}$ 
using the two X-ray detectors FPMA and FPMB (ObsID: 80001042002), 
simultaneously with the {\rm XMM-Newton} observation. 
The raw data were reprocessed using the {\rm NuSTAR} Data Analysis Software (NuSTARDAS, v2.1.2 included in HEASOFT V.6.31.1) and the calibration database (CALDB) version 20230420. We used the \texttt{nuproducts} task to generate barycenter-corrected light curves and spectra. 
We extracted the source data from a circular region with a radius of $\ang{;;49}$ and the background from an annulus with an outer radius 
of $\ang{;;98}$ surrounding the source region.

\section{Analysis and Results}\label{sec:analysis}
\subsection{Timing Analysis}

\begin{figure*}[t]
    \gridline{
    \fig{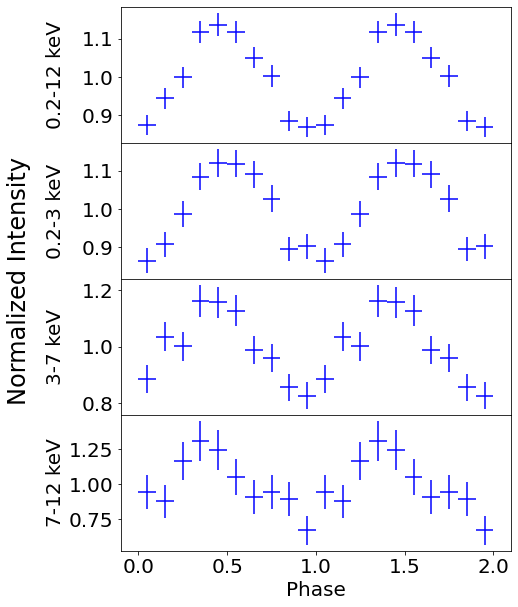}{0.345\textwidth}{(a) {\rm XMM-Newton} EPIC-pn}
    \fig{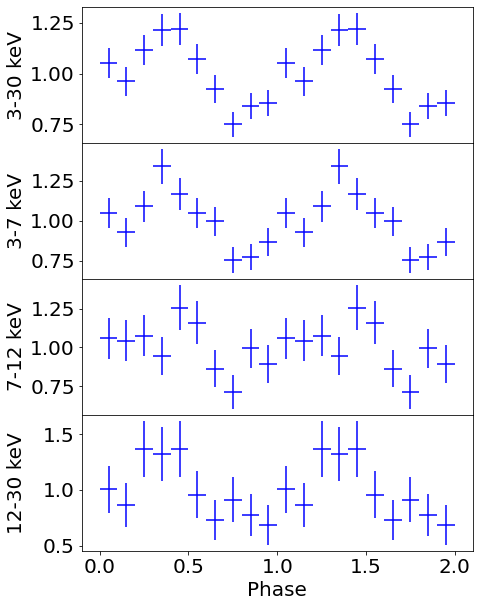}{0.32\textwidth}{(b) {\rm NuSTAR} FPMA}
    \fig{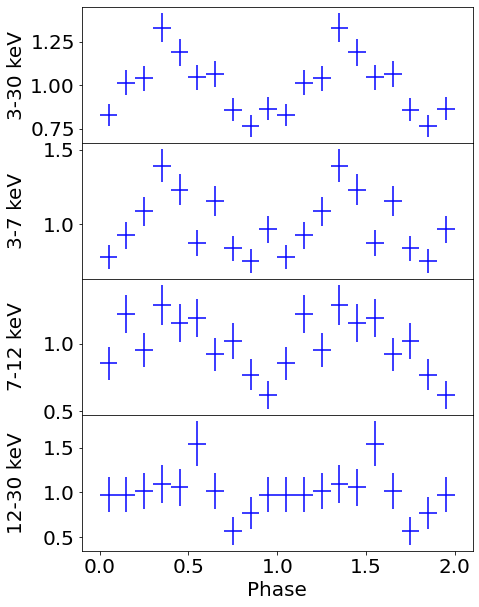}{0.32\textwidth}{(c) {\rm NuSTAR} FPMB}
    }
    \caption{Pulse profiles from the individual instruments with different energy ranges. For (a), the energy ranges are as follows (from the top to the bottom): $0.2-\SI{12}{keV},~0.2-\SI{3}{keV},~3-\SI{7}{keV}$ and $7-\SI{12}{keV}$. For (b) and (c), the energy ranges are as follows in the same order: $3-\SI{30}{keV},~3-\SI{7}{keV},~7-\SI{12}{keV}$ and $12-\SI{30}{keV}$. The profiles are normalized by the average count rate over the observations.}
    \label{fig:pulse_profile}
\end{figure*}

We perform a pulsation search using the EPIC-pn, FPMA, and FPMB light curves. Only 3--10\,keV events are used for this analysis, because all the three detectors have sensitivity to this energy band. 
We apply an accelerated Fourier method, which searches a grid in the frequency $f$ and frequency-derivative $\dot{f}$ space. 
To do this, we introduce the \texttt{HENaccelsearch} task in the HENDRICS software package \citep{bachetti_maltpynt_2015}. 
Subsequently, the best combination of $(f,\dot{f})$ obtained from this search (for each detector) is analyzed in more detail using the $Z_n^2$ statistics \citep{buccheri_search_1983} available in the \texttt{HENzsearch} task. 
The resulting period and its derivative are as follows: 
$P = \SI{1.13743+-0.00002}{s}$ and $\dot{P}=-5.38^{+1.02}_{-0.92}\times10^{-9}~\si{s.s^{-1}}$ at 56847.9 (MJD) for the EPIC-pn, 
$P = \SI{1.13766+-0.00001}{s}$ and $\dot{P}=-5.18^{+0.18}_{-0.17}\times10^{-9}~\si{s.s^{-1}}$ at 56847.4 (MJD) for the FPMA, 
and $P = \SI{1.13766+-0.00001}{s}$ and $\dot{P}=-5.20^{+0.18}_{-0.16}~\times10^{-9}~\si{s.s^{-1}}$ at 56847.4 (MJD) for the FPMB. 
The obtained values of $P$ and $\dot{P}$ are consistent among the three detectors when the reference epoch is aligned, and also agree with the previously reported values of \cite{israel_accreting_2017}. 
Since the most stringent constraints on $P$ and $\dot{P}$ are obtained for the FPMA, we use their mean values to fold the light curves of all the three instruments.

Figure \ref{fig:pulse_profile} shows the resulting folded light curves in different energies, obtained from the (a) EPIC-pn, (b) FPMA, and (c) FPMB. 
In the profiles for the whole energy range
(i.e., 0.2--12\,keV for the EPIC-pn and 3--30\,keV for the FPMA/FPMB),
the peak and minimum of the pulse are found at phase 0.4--0.5 and phase 0.9--1.1, respectively. The energy-dependent phase shift is clearly seen in the EPIC-pn profiles, confirming the previous results of \cite{israel_accreting_2017}. 
We also confirm the larger pulse fraction achieved at higher energies, suggesting that the hard X-ray emission predominantly originates from the accretion flow formed along the magnetic fields of the neutron star.

\subsection{Spectral Analysis}\label{subsec:spectral_analysis}

\begin{figure}
    \centering
    \includegraphics[scale=0.33]{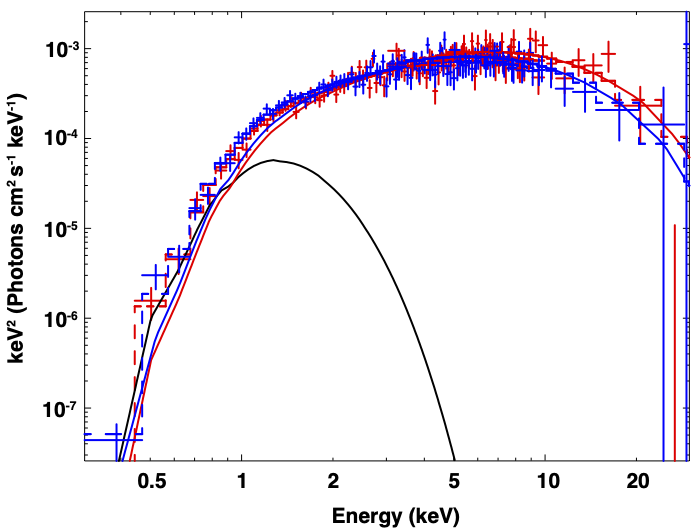}
    \includegraphics[scale=0.28]{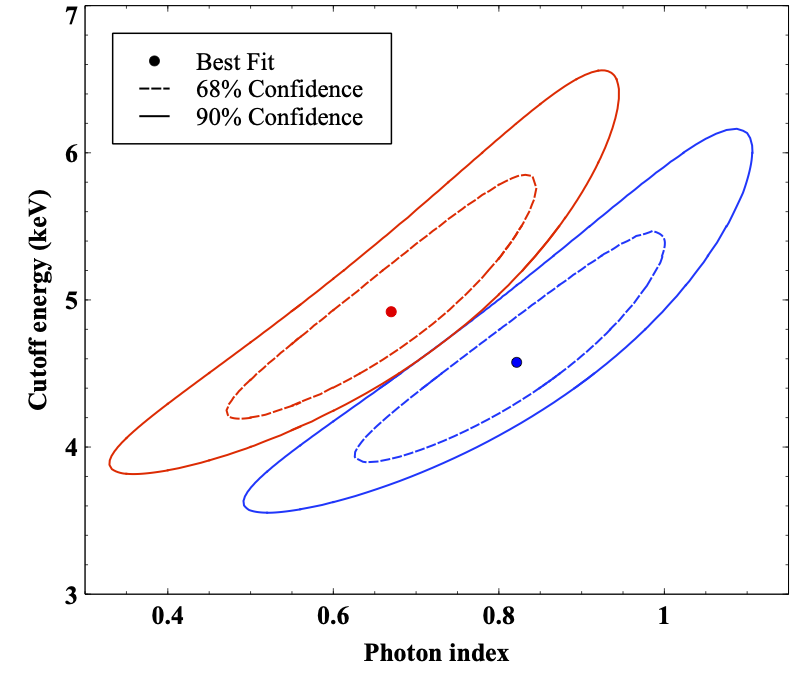}
    \caption{(Top) Response unfolded spectra of NGC 5907 ULX1. The pre-peak (phase $0.1\--0.3$) and post-peak (phase $0.6\--0.8$) data are shown in red and blue, respectively. The spectra are rebinned, and those from the different instruments share the same marker for visual clarity. For the same reason, the NuSTAR FPMB data are omitted from the plot. The lines represent the best-fit \texttt{diskpbb+cutoffpl} model components with $p = 0.5$ (see text). The colored dashed and solid lines correspond to the contribution of the total and \texttt{cutoffpl} models, respectively. The black solid line represents the \texttt{diskpbb} component (for the EPIC-pn only). (Bottom) Confidence contours of the photon index versus the cutoff energy of the \texttt{cutoffpl} model component for the case of $p=0.5$. Confidence levels from the pre-peak are shown in red, while those from the post-peak are in blue. The dashed and solid lines represent the 68\% and 90\% confidence levels, respectively. The circle in each color indicates the best-fit value. 
    }
    \label{fig:spectrum}
\end{figure}

\begin{table*}[t]
    \caption{Best-fit Model Parameters for Pre-peak and Post-peak Spectra\label{tab:parameter}}
    \begin{tabular}{llccccc} \hline\hline
    ~ & ~ & \multicolumn{2}{c}{$p=0.5$} & ~ & \multicolumn{2}{c}{$p=0.75$} \\ \cline{3-4}\cline{6-7}
    Component & Parameter & Pre-peak & Post-peak & ~ & Pre-peak & Post-peak\\ \hline
    \texttt{tbabs} & $N_\mathrm{H}$~$(\SI{e22}{cm^{-2}})$\tablenotemark{a} & \multicolumn{2}{c}{$0.85^{+0.14}_{-0.11}$} & ~ & \multicolumn{2}{c}{$0.79^{+0.14}_{-0.12}$} \\  
    \texttt{diskpbb} & $T_\mathrm{in}~(\si{keV})$ & \multicolumn{2}{c}{$0.31^{+0.07}_{-0.05}$} & ~ & \multicolumn{2}{c}{$0.30^{+0.06}_{-0.05}$} \\
    ~ & $N$\tablenotemark{a} & \multicolumn{2}{c}{$1.2^{+5.8}_{-1.1}$} & ~ & \multicolumn{2}{c}{$3.2^{+13.6}_{-2.6}$} \\
    \texttt{cutoffpl} & $\Gamma$\tablenotemark{a} & $0.67^{+0.18}_{-0.20}$ & $0.82^{+0.18}_{-0.19}$ & ~ & $0.66^{+0.19}_{-0.21}$ & $0.80^{+0.20}_{-0.21}$  \\
    ~ & $E_\mathrm{fold}~(\si{keV})\tablenotemark{a}$ & $4.9^{+0.9}_{-0.7}$ & $4.6^{+0.9}_{-0.7}$ & ~ & $4.9^{+1.0}_{-0.8}$ & $4.5^{+1.0}_{-0.7}$ \\
    ~ & $F~(\SI{e-12}{erg.cm^{-2}.s^{-1}})\tablenotemark{b}$ & $3.0\pm 0.1$ & $2.8\pm 0.1$ & ~ & $3.0\pm 0.1$ & $2.8\pm 0.1$ \\
    \texttt{constant} & $C_\mathrm{pn}$ & \multicolumn{2}{c}{$1$~(fixed)} & ~ & \multicolumn{2}{c}{$1$~(fixed)} \\
    ~ & $C_\mathrm{FPMA}$ & \multicolumn{2}{c}{$1.13\pm0.04$} & ~ & \multicolumn{2}{c}{$1.13^{+0.04}_{-0.03}$} \\
    ~ & $C_\mathrm{FPMB}$ & \multicolumn{2}{c}{$1.11^{+0.04}_{-0.03}$} & ~ & \multicolumn{2}{c}{$1.11\pm 0.04$} \\ \hline
    \multicolumn{2}{l}{$PG$-statistic/dof} & \multicolumn{2}{c}{$7426.10/9416$} & ~ & \multicolumn{2}{c}{$7426.43/9416$} \\ \hline
    \end{tabular}
    \tablecomments{}
    \tablenotetext{a}{Uncertainties are obtained using XSPEC \texttt{steppar} command in the 2D-spaces between $N_\mathrm{H}$ and $N$, and between $\Gamma$ and $E_\mathrm{fold}$.}
    \tablenotetext{b}{Flux in the $0.3$--$\SI{40}{keV}$.}
\end{table*}

To perform phase-resolved spectroscopy, we divide the pulse period into 10 phase bins with equal width, and extract spectra from each bin. 
We then merge the spectra from phases 0.1--0.2 and 0.2--0.3 to represent the pre-peak part and phases 0.6--0.7 and 0.7--0.8 to represent the post-peak part of the pulse cycle. 
The luminosities at these two phases (0.1--0.3 and 0.6--0.8) are 
nearly identical to each other. 
We use the EPIC-pn data in the 0.3--10\,keV band and the FPMA/FPMB data in the 3--30\,keV band for the spectral analysis. 
The unfolded spectra of the pre-peak and post-peak phases are shown in the top panel of Figure \ref{fig:spectrum} (red and blue, respectively),
confirming the harder spectrum in the former, as implied in the pulse profiles (Figure \ref{fig:pulse_profile}). 
In Figure~\ref{fig:appendix_spectrum} in Appendix~\ref{sec:appendix}, we also show the spectra extracted from the eight different phases (i.e., the pre-peak, post-peak, and other 6 phases with the 0.1 phase width). 

For more quantitative studies, we fit the spectra with a model consisting of 
a \texttt{diskpbb} component and an power law with an exponential cutoff (\texttt{cutoffpl})   
using the XSPEC analysis package version v12.13.0c \citep{arnaud_xspec_1996}. 
The \texttt{diskpbb} model represents multicolor blackbody emission from 
an optically-thick accretion disk with a radial temperature gradient \citep{mineshige_time-dependent_1994}. 
The free parameters of this component are the innermost temperature ($T_{\rm in}$), 
the index of the radial dependence of the disk temperature ($p$, where $T(r) \propto r^{-p}$), 
and the normalization ($N$). 
Theoretically, the index $p$ is expected to range from 0.5 to 0.75, 
depending on the mass accretion rate and the resulting physical properties of the disk. 
When the accretion rate is below the Eddington limit, a geometrically-thin, 
standard disk is formed due to efficient radiative cooling, 
and $p = 0.75$ is expected accordingly \citep{shakura_black_1973}. 
When the accretion rate is high, on the other hand, the disk increases its geometrical thickness 
due to strong radiation pressure and advection, leading to a flatter temperature gradient ($p = 0.5$) achieved within the radius called ``spherization radius'' 
(e.g., \citealt{poutanen_supercritically_2007}). 
This type of disk is commonly referred to as ``slim disk'' \citep{jaroszynski_supercritical_1980,abramowicz_slim_1988,watarai_galactic_2000,watarai_slim-disk_2001}.
The other model component, \texttt{cutoffpl}, is responsible for the pulsed emission associated with the accretion flow rotating with the magnetosphere of the neutron star. 
This phenomenological model is often introduced to fit typical ULXP spectra (e.g., \citealt{brightman_spectral_2016,walton_super-eddington_2018}), 
although the radiation mechanism related to this component is still under debate (see \S4.3).
The free parameters are the photon index ($\Gamma$), the e-folding energy of exponential cutoff ($E_{\rm fold}$), 
and flux in the 0.3--40\,keV band ($F$).

In addition, we introduce the Tuebingen-Boulder interstellar medium absorption model, \texttt{tbabs}, 
to account for the foreground absorption, assuming the solar abundances of \cite{wilms_absorption_2000}. 
The hydrogen column density ($N_{\rm H}$) for the Galactic absorption is fixed to $\SI{1.38e20}{cm^{-2}}$ \citep{kalberla_leidenargentinebonn_2005}, whereas the value for the absorption in the NGC 5907 galaxy is 
left as a free parameter. 
We assume that the absorption intrinsic to the source binary system is negligibly small compared to that in its host galaxy, 
since this galaxy is edge-on. 
Lastly, we introduce a cross-normalization parameter, \texttt{const}, to account for the systematic uncertainty in the effective area calibration among the instruments. The \texttt{const} value for the EPIC-pn ($C_\mathrm{pn}$) is fixed to unity, whereas 
those for the FPMA ($C_\mathrm{FPMA}$) and FPMB ($C_\mathrm{FPMB}$) are allowed to vary freely. 

With the model described above, we fit the eight phase-resolved spectra given in Appendix \ref{sec:appendix} 
simultaneously, assuming Poisson data with Gaussian background \citep[PG-statistic:][]{cash_parameter_1979}. 
Since the emission from the accretion disk is considered to be phase-invariant, the parameters of the \texttt{diskpbb} component is tied among the eight spectra. The parameters of the \texttt{cutoffpl} components, responsible for the pulsed emission, are fitted independently.
We then obtain a reasonably good fit with the $PG$-stat value of 7426.11 (for 9415 dof), 
but the parameter $p$ is not constrained within the allowable range of 0.5--0.75. 
Therefore, we repeat the fitting by fixing the $p$ value to 0.5 or 0.75. 
The two extreme cases result in similarly good fits with $PG$-stat values of 7426.10 and 7426.43 for $p=0.5$ and $p=0.75$, respectively (with dof of 9416 for both). 
The best-fit results are given in Table \ref{tab:parameter}. 
No significant difference is found between the results with $p=0.5$ and $p=0.75$; all the best-fit values are consistent within the statistical errors. 

We confirm that the \texttt{cutoffpl} component is slightly harder in the pre-peak phase than in the post-peak phase. 
Since the parameters $\Gamma$ and $E_\mathrm{fold}$ are tightly correlated with each other, we investigate their statistical errors in the 2-dimensional space. The bottom panel of Figure \ref{fig:spectrum} shows the confidence contours of these parameters for the slim disk ($p=0.5$) case. 
A significant difference is found between the two phases at the $90~\%$ confidence level. A similar result is obtained for the $p=0.75$ case as well.


\section{Discussion}\label{sec:discussion}

\subsection{Summary of the Results and Comparison with Preceding Work}
We have performed phase-resolved spectral analysis of the broadband X-ray spectra of NGC 5907 ULX1 in its bright phase in 2014. 
The spectra are well-described by the model consisting of a time-invariant accretion disk component and a pulsed emission component with foreground extinction. The disk component can be equally well described by a geometrically-thin (standard) disk and a geometrically-thick (slim) disk, 
although the latter would be favored from the nature of the ULX.
The pulsed emission component, associated with the corotating accretion flow onto the neutron star within the magnetosphere, can be reasonably modeled by a power law with an exponential cutoff.
Its spectral shape, characterized by the combination of the photon index and cutoff energy, is different between the pre-peak and post-peak phases; the former spectrum is harder than the latter one. 

The spectra we have presented in this paper were studied in some previous work as well. 
\cite{furst_spectral_2017} analyzed phase-averaged spectra for the purpose of investigating the spectral changes as a function of the 78-day super-orbital phase of the binary system.
They first attempted a model similar to ours (i.e., multicolor disk blackbody + cutoff power law), and obtained the best-fit innermost disk temperature consistent with our results (i.e., $T_{\rm in} \sim 0.3$\,keV). 
However, this model left systematic residuals above $\sim$\,10\,keV, unlike our results based on the phase-resolved spectroscopy. 
In their work, this hard excess was attributed to the Comptonization of disk photons, and a simple Comptonization model  \citep[\texttt{simpl}:][]{steiner_simple_2009} was introduced as a convolution component applied to the \texttt{diskpbb} component.
As a result, they obtained an innermost disk temperature ($T_{\rm in}$) of $\sim$\,3\,keV, significantly higher than our result. 
We note that the \texttt{simpl} model in XSPEC is an empirical model of Comptonization that converts a fraction of input seed photons into higher-energy photons with a power law distribution. 
In this model, the spatial distribution of the Comptonizing electrons is assumed to be uniform around the seed photon emitting regions, so all the seed photons are scattered at the same probability \citep{steiner_simple_2009}. 
This means that the \texttt{simpl*diskpbb} model introduced by \cite{furst_spectral_2017} represents the situation where the accretion disk is uniformly surrounded by the Comptonization region. 
Given the presence of the pulsation, we believe that the hard X-ray photons originate from the region much closer to the neutron star, and thus this empirical model is not introduced in our analysis. 




\cite{walton_evidence_2018} conducted phase-resolved spectroscopy, but the phase dependence in the spectral shape was not taken into account. In fact, they extracted spectra from the brightest and faintest quarters of the pulse cycle, and subtracted the latter from the former (i.e., pulse-on -- pulse-off) to determine the spectral shape of the pulsed emission component using a \texttt{cutoffpl} model. Subsequently, the phase-resolved spectra were fitted with the \texttt{diskpbb} and \texttt{cutoffpl} components, similar to our analysis, but
the photon index and cutoff energy of the pulsed component were fixed to the values derived from the analysis of  ``pulse-on -- pulse-off'' spectrum. 
The innermost disk temperature was then obtained to be $\sim$\,3\,keV, similar to the result of \cite{furst_spectral_2017} based on the \texttt{simpl * diskpbb} model. 
In our analysis, we have taken into account the spectral shape variation among the phase bins, obtaining a substantially lower $T_{\rm in}$ value ($\sim$\,0.3\,keV). Moreover, no systematic residuals are found in the hard X-ray band over the model consisting of the \texttt{diskpbb} plus \texttt{cutoffpl} components. Keeping them in mind, we discuss the implication of our results in the following subsections.

\subsection{Disk Inclination Angle}\label{sec:Inclination}

As discussed above, the disk emission is well reproduced by either the standard or slim disk model in our spectral analysis. 
Its bolometric luminosity is obtained to be 
$L_\mathrm{standard} = (1.9 \pm 0.1)\times 10^{40}~\si{erg.s^{-1}}$ for the standard disk case and $L_\mathrm{slim} = (1.7 \pm 0.1)\times 10^{41}~\si{erg.s^{-1}}$ for the slim disk case, assuming the distance to the source to be $\SI{17.1}{Mpc}$ \citep{tully_cosmicflows-3_2016}. 
Given the extremely high luminosity observed in this source, 
it is reasonable to assume that the innermost disk is geometrically-thick due to the super-critical accretion, but a geometrically-thin part should also be present in the regions outside the spherization radius.
Therefore, the actual luminosity of the whole accretion disk is considered to be in between $L_\mathrm{standard}$ and $L_\mathrm{slim}$. 
On the other hand, the luminosity of the power-law component is 
obtained to be $L_\mathrm{pow} = \SI{1.0e41}{erg.s^{-1}}$, 
regardless of the applied disk model. The uncertainty to this value is negligibly small.
The bolometric luminosity of the whole system is, therefore, constrained to be
\begin{equation}
	1.2 \leq L_{\rm bol}~[10^{41}~\si{erg.s^{-1}}] \leq 2.8~.
	\label{bolometric_luminosity}
\end{equation}

The accretion disk is truncated at the magnetospheric radius $R_\mathrm{M}$, where the magnetic field pressure of the neutron star becomes comparable to the ram pressure of the accreting matters. 
Following Equation~S6 of \cite{israel_accreting_2017} and assuming that the geometry of the innermost accretion disk is nearly spherical, we estimate this radius to be 
\begin{align}
    R_\mathrm{M} = &~\num{6.6e2}~\qty(\frac{L_{\rm bol}}{10^{41}~\si{erg.s^{-1}}})^{-2/7}\qty(\frac{B_{\rm NS}}{10^{13}~\si{G}})^{4/7} \nonumber \\
    &~~\times\qty(\frac{M_\mathrm{NS}}{1.4~M_\odot})^{1/7}\qty(\frac{R_\mathrm{NS}}{10^6~\si{cm}})^{10/7}~\si{km}, \label{magnetospheric_radius}
\end{align}
where $B_{\rm NS}$, $M_\mathrm{NS}$, and $R_\mathrm{NS}$ are the surface polar magnetic field strength, mass, and radius of the neutron star, respectively. 

Note that this equation is based on the analytical model originally developed by \cite{lamb_model_1973},
where a moderate accretion rate is assumed. 
It is therefore not obvious if this equation is valid for the system with a super-critical accretion rate. 
Nevertheless, we first estimate the disk inclination angle based on this model for a direct comparison with the previous work. 



Given that NGC 5907 ULX1 was in its bright phase at the time of the observations, $R_\mathrm{M}$ is considered to be smaller than the corotation radius $R_\mathrm{c}$, where the Keplerian orbital angular velocity is equal to the angular velocity of the neutron star rotation. Otherwise, the centrifugal force prevents the accretion flow so that the system becomes X-ray dim. 
Using the spin period of the neutron star $P$, we derive the corotation radius as: 
\begin{equation}
    R_\mathrm{c} = \qty(\frac{GM_\mathrm{NS}P^2}{4\pi^2})^{1/3}, \label{corotation_radius}
\end{equation}
where $G$ is the gravitational constant. 
Since $P = \SI{1.137}{s}$ for NGC 5907 ULX1 at the time of the observations, 
Eq.\,\ref{corotation_radius} gives $R_\mathrm{c} = \SI{1.8e3}{km}$, 
when $M_{\rm NS} = 1.4\,M_\odot$ is assumed. 
Therefore, the condition of $R_\mathrm{M} < R_\mathrm{c}$ leads to the relationship between $L_{\rm bol}$ and $B_{\rm NS}$ as:
\begin{equation}
	L_{\rm bol} > 2.8 \times 10^{39} \qty(\frac{B_{\rm NS}}{10^{13}~\si{G}})^2~{\rm erg\,s}^{-1}.
 ~\label{luminosity_ll}
\end{equation}
This gives a more stringent lower limit on the luminosity than that in Equation~\ref{bolometric_luminosity} when $B_{\rm NS} > 6.5\times 10^{13}~\si{G}$. 
If $B_{\rm NS}$ exceeds $9.9 \times 10^{13}~\si{G}$, 
however, there is no luminosity range that satisfies 
Equations~\ref{bolometric_luminosity} and \ref{luminosity_ll} simultaneously. 
Therefore, this value is regarded as the upper limit of the magnetic field strength 
at the neutron star surface. 

\begin{figure*}
    \begin{minipage}{0.66\columnwidth}
        \centering
        \includegraphics[scale=0.24]{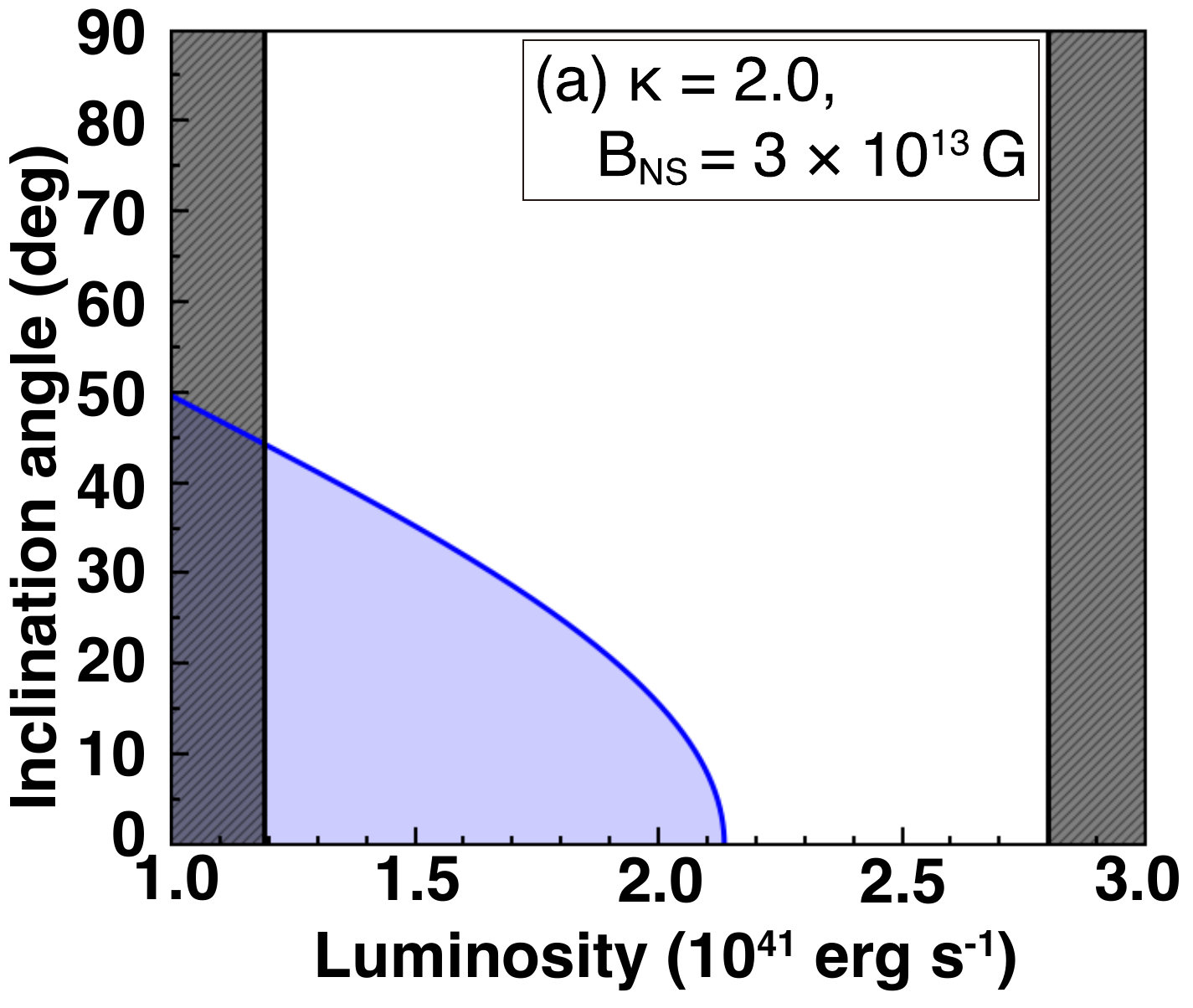}
    \end{minipage}
    \begin{minipage}{0.66\columnwidth}
        \centering
        \includegraphics[scale=0.24]{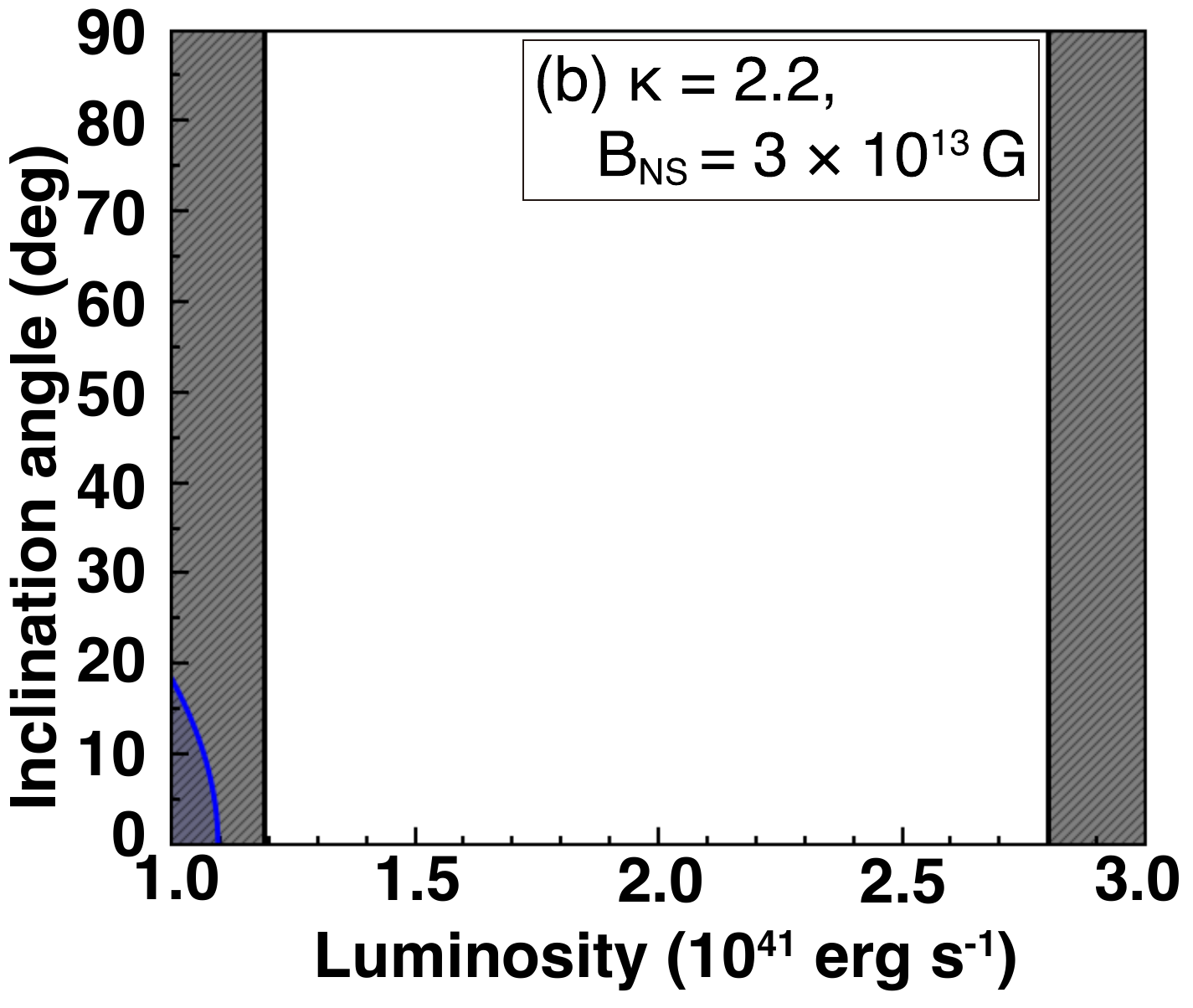}
    \end{minipage}
    \begin{minipage}{0.66\columnwidth}
        \centering
        \includegraphics[scale=0.24]{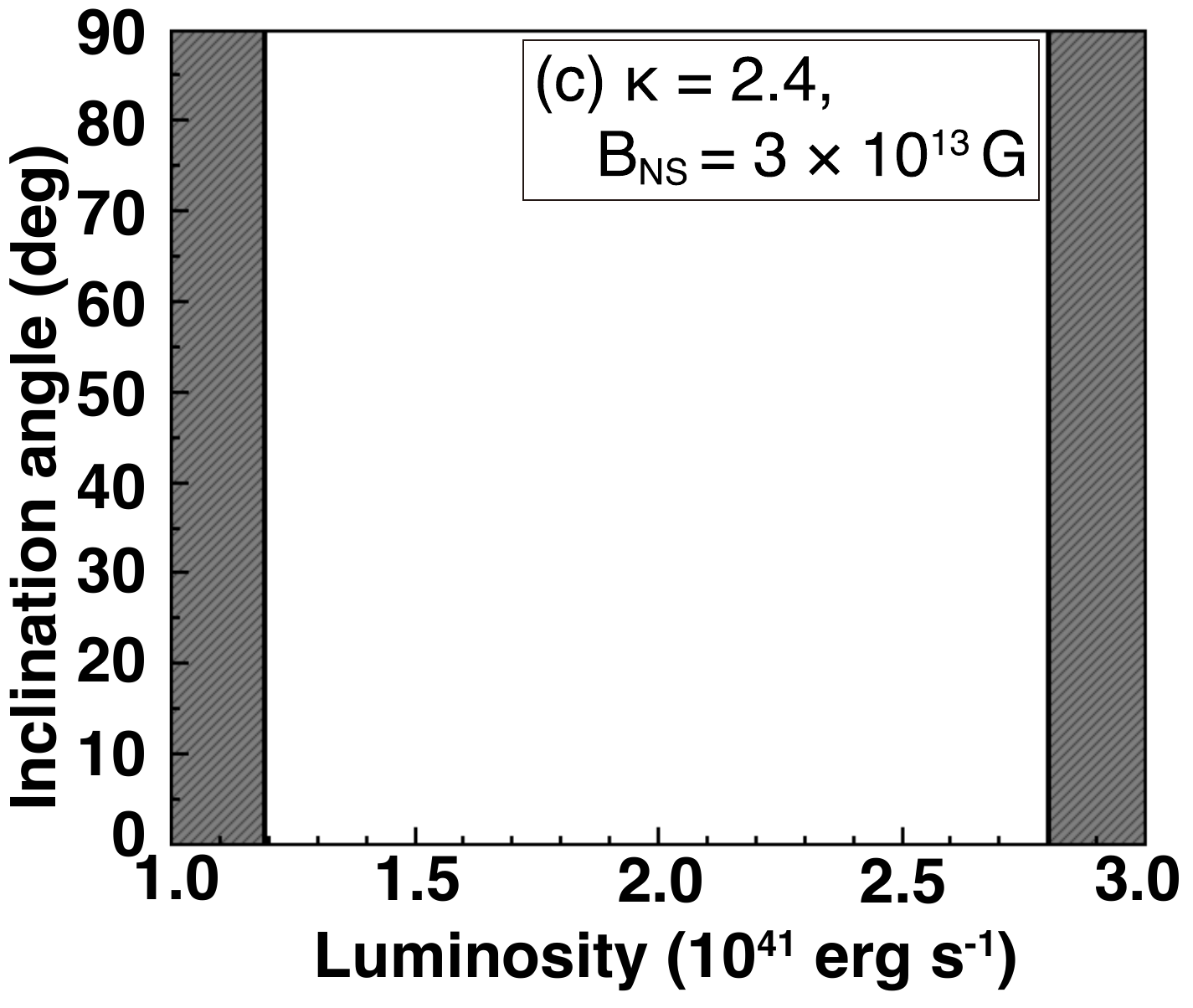}
    \end{minipage}\\
    \begin{minipage}{0.66\columnwidth}
        \centering
        \includegraphics[scale=0.24]{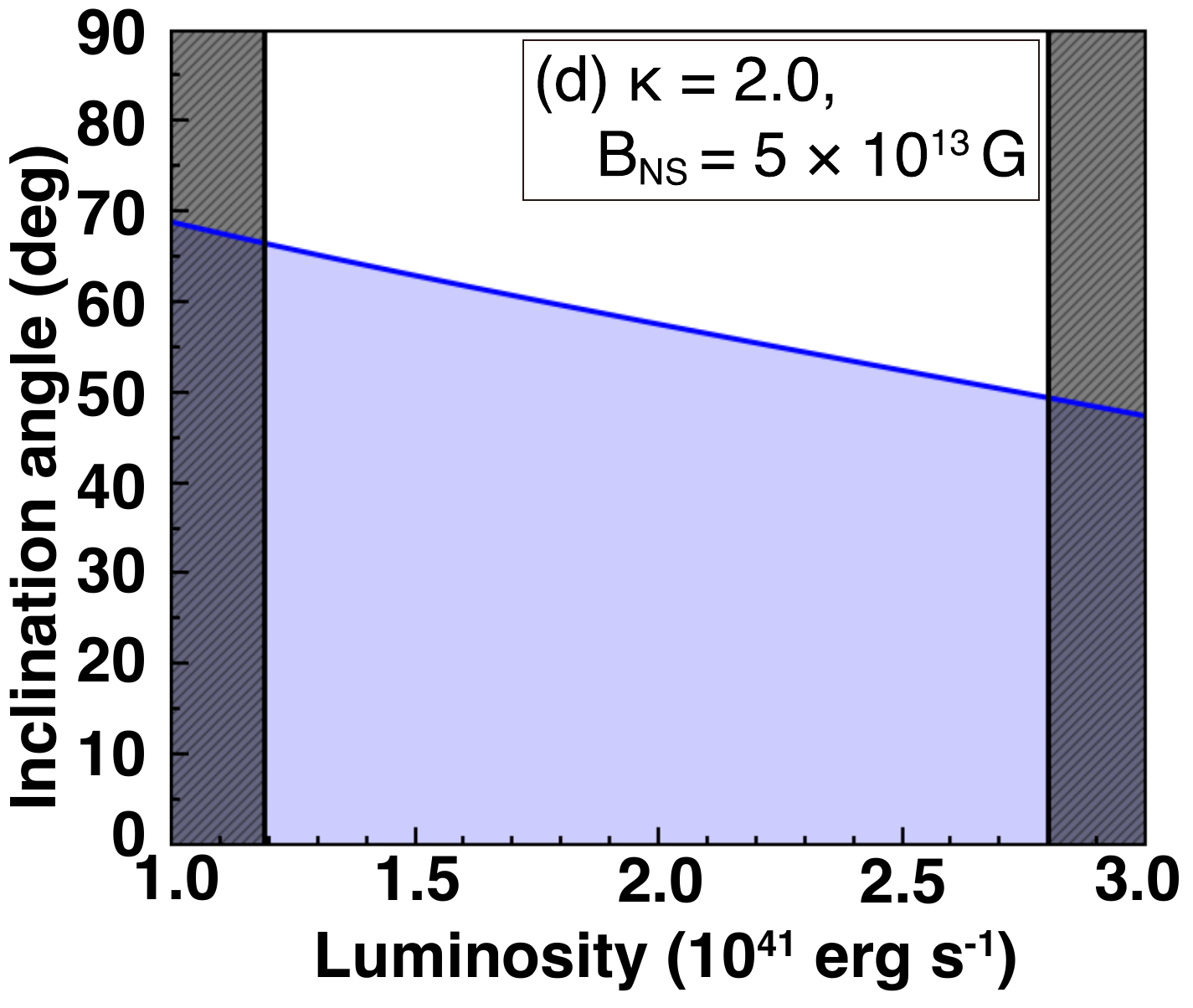}
    \end{minipage}
    \begin{minipage}{0.66\columnwidth}
        \centering
        \includegraphics[scale=0.24]{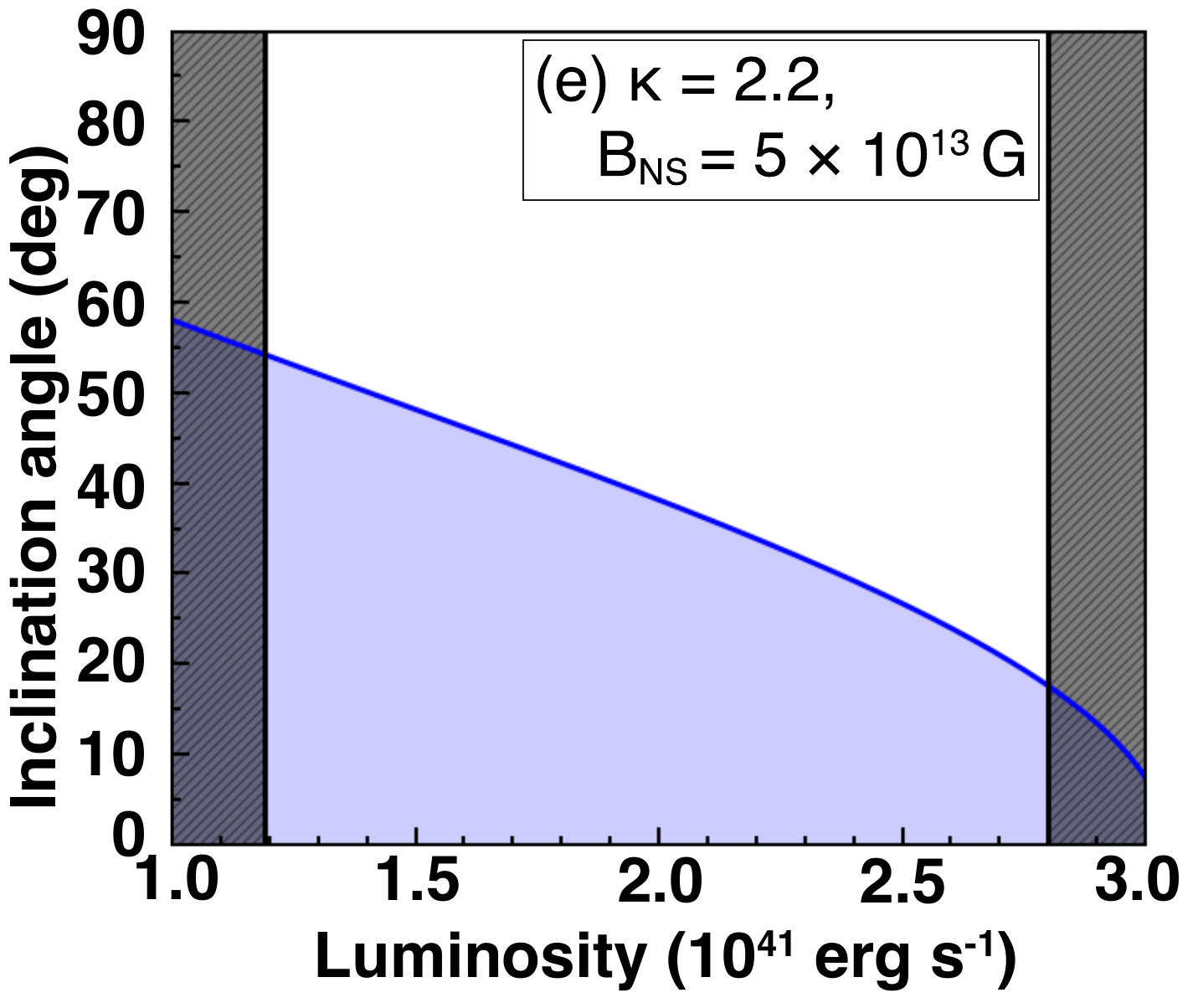}
    \end{minipage}
    \begin{minipage}{0.66\columnwidth}
        \centering
        \includegraphics[scale=0.24]{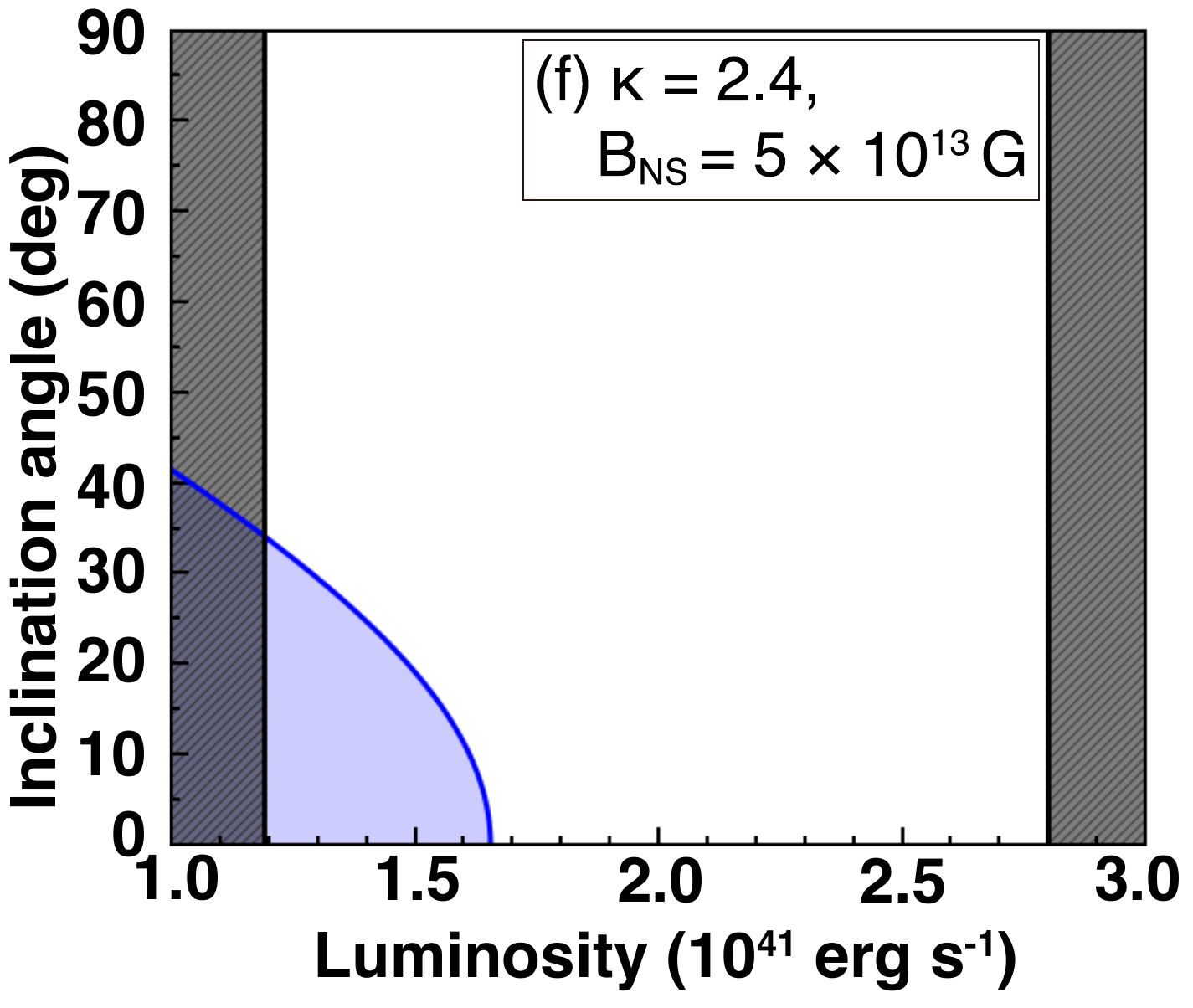}
    \end{minipage}\\
    \begin{minipage}{0.66\columnwidth}
        \centering
        \includegraphics[scale=0.24]{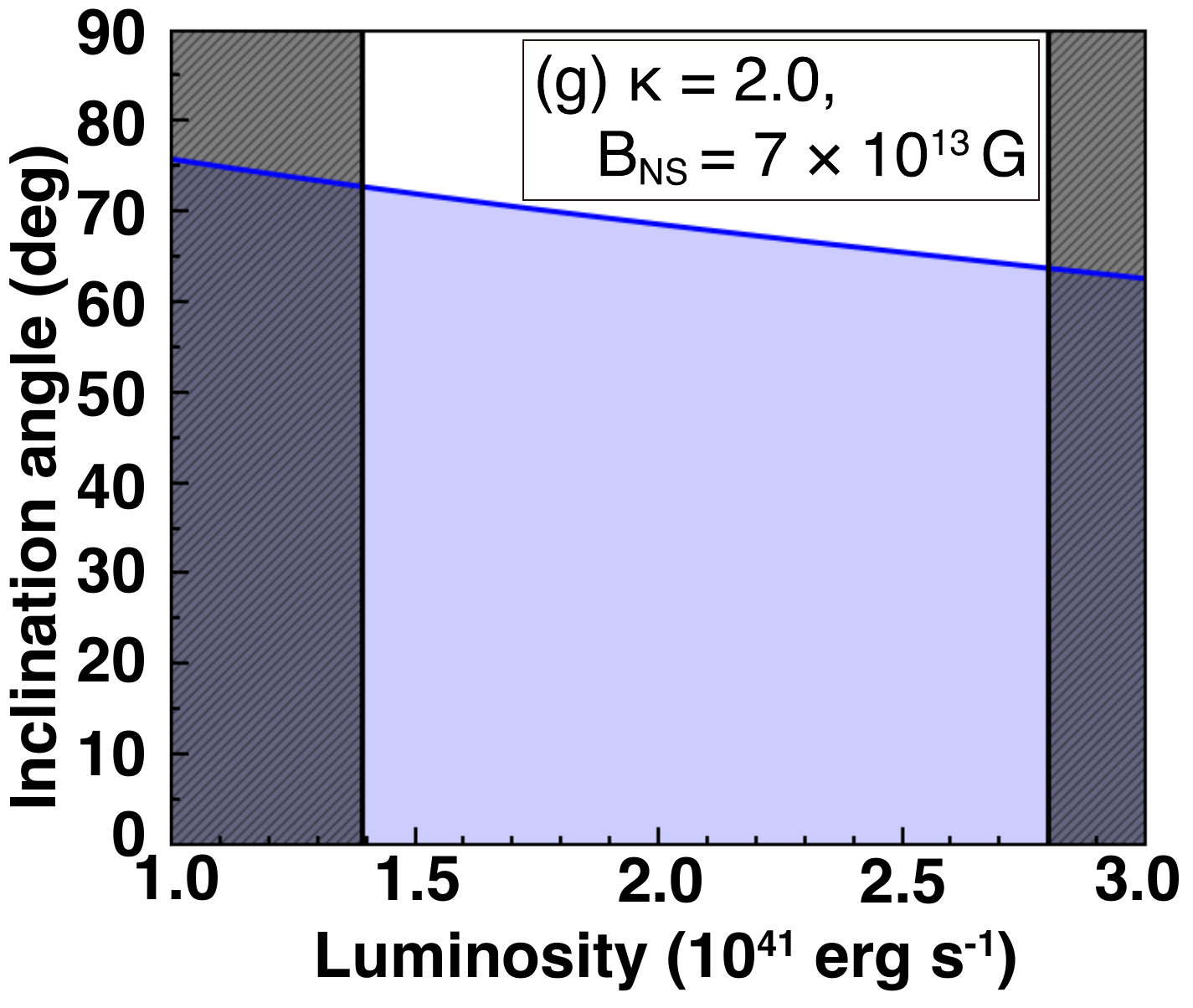}
    \end{minipage}
    \begin{minipage}{0.66\columnwidth}
        \centering
        \includegraphics[scale=0.24]{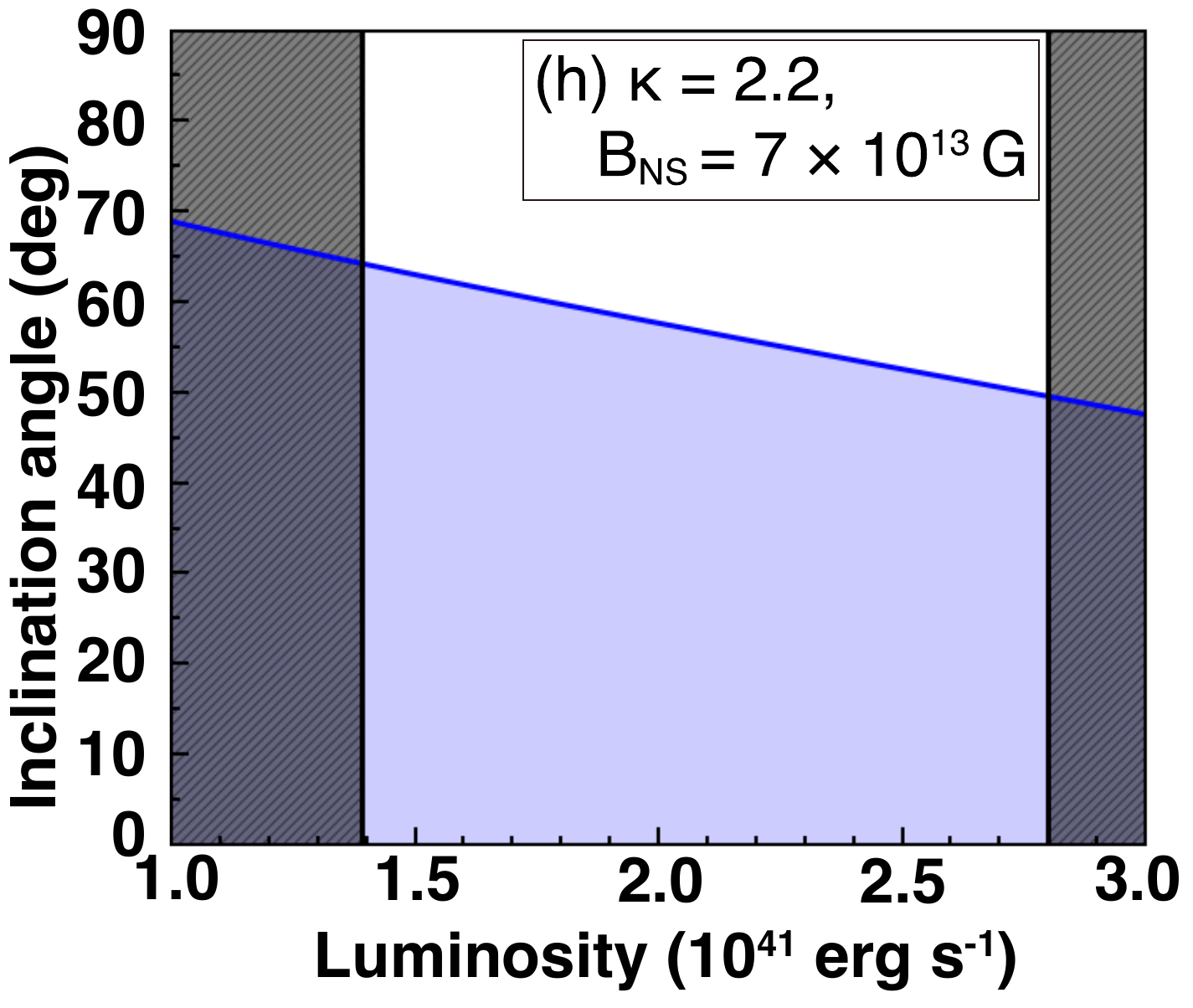}
    \end{minipage}
    \begin{minipage}{0.66\columnwidth}
        \centering
        \includegraphics[scale=0.24]{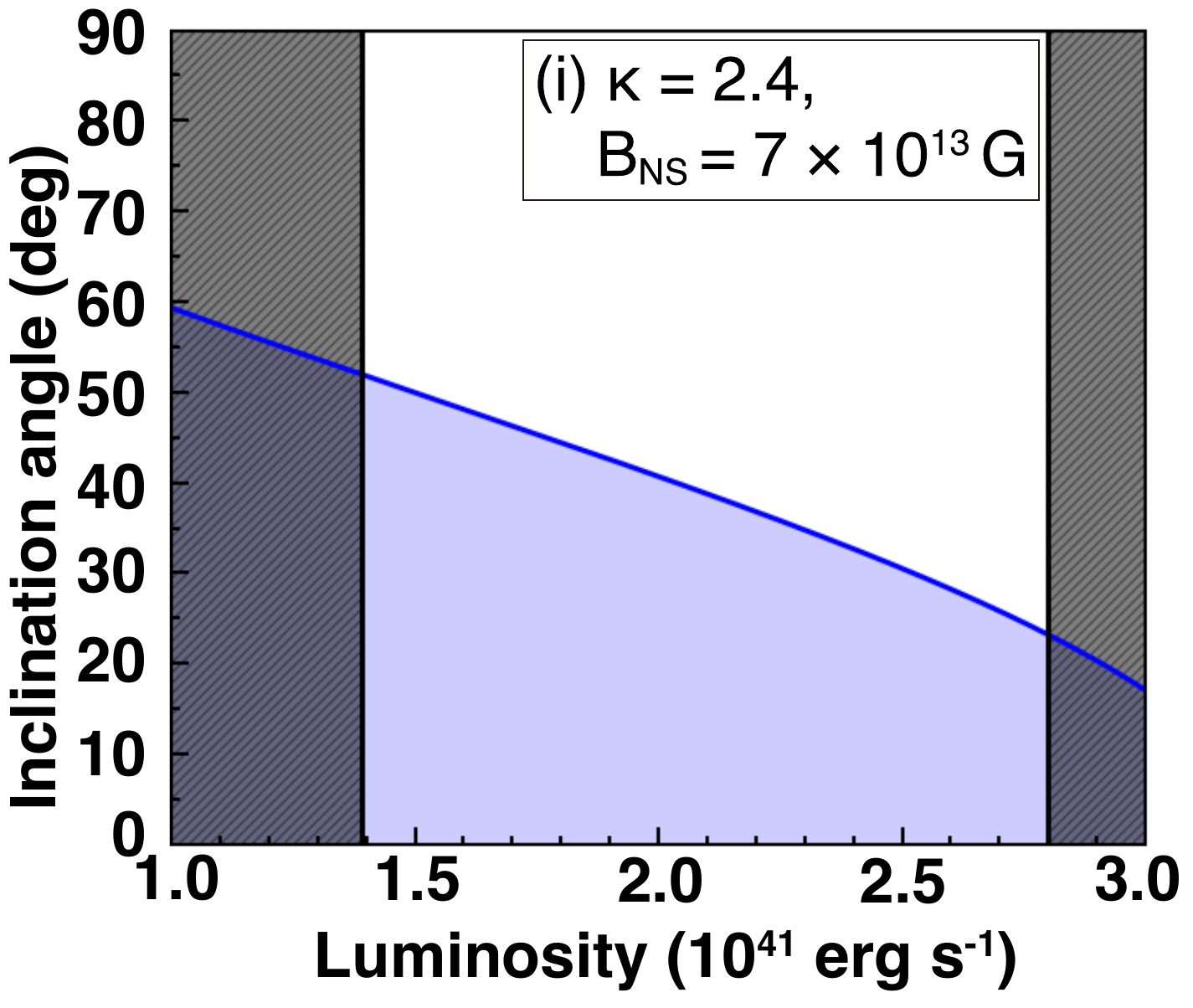}
    \end{minipage}
    \caption{The inclination angle of the accretion disk as a function of the bolometric luminosity under different assumptions of the color-hardening factor $\kappa$ ($[2.0, 2.2, 2.4]$) and magnetic field $B_{\rm NS}$ ($[3.0, 5.0, 7.0]\times 10^{13}~\si{G}$). 
    The blue regions indicate the allowed range of the inclination angle with 1$\sigma$ confidence level derived from Equation \ref{cos}.
    The gray areas represent where the luminosity is outside the limits. \label{fig:lum_inc}}
\end{figure*}

Since the \texttt{diskpbb} model is an approximation of a multi-color accretion disk, its normalization ($N$) provides an estimate of the inner-disk radius ($R_\mathrm{in}$) as:  
\begin{equation}
    R_\mathrm{in} = \qty(\frac{N}{\cos\theta})^{\!1/2} d_{10}\,\epsilon\,\kappa^2
    \label{r_in}
\end{equation}
(e.g., \citealt{makishima_nature_2000}), where $d_{10},~\epsilon,~\kappa$ and $\theta$ are the distance to the object in units of $\SI{10}{kpc}$, geometric factor, 
color-hardening factor, and the inclination angle of the disk, respectively. 
The geometric factor $\epsilon$ compensates for the difference in the innermost edge boundary condition between the assumption in the \texttt{diskpbb} model and the actual accretion disk. 
In the case of the standard disk, $\epsilon = 0.412$ is often applied \citep{kubota_evidence_1998}. 
On the other hand, \cite{vierdayanti_black_2008} argued that $\epsilon = 0.353$ is more appropriate for the super-critical accretion case, and \cite{soria_slim-disk_2015} indeed used this value to estimate the black hole mass of M83 ULX-1. 
It is more difficult to estimate an appropriate $\epsilon$ value for a magnetized neutron star, because the accretion disk is truncated due to the presence of the magnetosphere. 
Nonetheless, we assume $\epsilon = 0.353$ in the following calculations.  
The color-hardening factor $\kappa$ is introduced to convert the color temperature determined from the spectral fit into the effective temperature of the disk. 
At sub-Eddington accretion rates, $\kappa \approx 1.7$ is known to be 
a good approximation \citep{shimura_spectral_1995,davis_relativistic_2005}. 
However, the value can be as high as 2.0--3.0 at higher accretion rates 
(e.g., \citealt{watarai_galactic_2000,isobe_suzaku_2012}). 

As already mentioned, the accretion disk is thought to be truncated at $R_\mathrm{M}$, allowing us to assume $R_\mathrm{in} = R_\mathrm{M}$. 
Therefore, Equations~\ref{magnetospheric_radius} and \ref{r_in} and the best-fit $N$ value give a constraint on the disk inclination angle as: 
\begin{align}
    \cos\theta = 2.6^{+12}_{-2.2} ~ \qty(\frac{\kappa}{2.0})^4 &\qty(\frac{B_{\rm NS}}{\SI{5e13}{G}})^{-8/7} \nonumber \\
    &\times\qty(\frac{L_\mathrm{bol}}{10^{41}~\si{erg.s^{-1}}})^{4/7}. 
    \label{cos}
\end{align}
Figure \ref{fig:lum_inc} illustrates the allowed inclination angle (in blue regions) as a function of bolometric luminosity for different values of $\kappa$ and $B_{\rm NS}$. 
The luminosity ranges that are ruled out from Equations~\ref{bolometric_luminosity} and 
\ref{luminosity_ll} are indicated as the gray regions. 
We find that a relatively low inclination angle is favored in general, 
and that a narrower allowable range is obtained for higher $\kappa$ 
and/or lower $B_{\rm NS}$. 
It should also be noted that, for $\kappa = 3$, 
the mathematically obvious requirement of $\cos\theta \leq 1$ 
is not satisfied with any combinations of $B_{\rm NS}$ and $L_{\rm bol}$ that are allowed in Equation~\ref{luminosity_ll}. 
We conclude, therefore, that moderate values (2.0--2.5) are favored for the color-hardening factor, despite the high accretion rate achieved in this system. 

Our result is in contrast to the previous study by \cite{furst_spectral_2017}, where a higher inclination angle of 
$\sim$\,$89^{\circ}$ was claimed. 
The origin of this discrepancy can be explained as follows. 
In the previous study, the innermost disk temperature $T_{\rm in}$ was obtained to be $\sim$\,3\,keV, significantly higher than our measurement ($\sim$\,0.3\,keV). 
Since the disk luminosity is proportional to $T_{\rm in}^4$, a lower normalization value ($N \sim\num{1e-3}$; see Table~1 for comparison) was derived to explain the observed X-ray flux. 
Consequently, the nearly maximum inclination angle (hence cos\,$\theta \approx 0$) was obtained from the equation of $R_{\rm M}$ and $R_{\rm in}$ (see Equations~\ref{magnetospheric_radius} and \ref{r_in}).
Our result can more naturally explain the detection of the pulsation, because the pulsating accretion flow can hardly be seen in the edge-on view due to the obstruction by the geometrically-thick disk that should be present in the super-critical accretion system. 
It is also worth noting that no evidence for an eclipse has been detected from this source, which also implies the low-inclination view of the system. 


Lastly, we introduce a model developed more recently by \cite{gao_magnetic_2021} to derive similar constraints on the disk inclination angle. 
This model was constructed by expanding the magnetic threaded disk models of 
\cite{ghosh_accretion2_1979,ghosh_accretion3_1979} and \cite{wang_disc_1987,wang_torque_1995} 
to the case of super-critical accretion, for the purpose of estimating the magnetic field of several observed ULXPs, including NGC\,5907 ULX1.
We follow Equation~17 of \cite{gao_magnetic_2021} to obtain the magnetospheric radius as: 

\begin{align}
    R_{\rm M} &= \xi^{7/9}\qty(\frac{B_{\rm NS}^4R_{\rm NS}^{12}}{2GM_{\rm NS}})^{1/9}\qty(\frac{2L_{\rm Edd}}{3GM_{\rm NS}})^{-2/9}  \nonumber \\
    &=\num{1.1e3}\qty(\frac{\xi}{0.7})^{7/9}\qty(\frac{B_{\rm NS}}{10^{13}~\si{G}})^{4/9} \nonumber \\
    &~~~~~~\times\qty(\frac{M_{\rm NS}}{1.4~M_\odot})^{1/9}\qty(\frac{R_{\rm NS}}{10^6~\si{cm}})^{4/3}~\si{km},
    \label{r_m_edd}
\end{align}

where $\xi$ is a factor expressing the magnetospheric radius as a fraction of the Alfv\'{e}n radius, and $L_{\rm Edd}$ is the Eddington luminosity. 
A remarkable difference from Equation~\ref{magnetospheric_radius} is that the bolometric luminosity $L_{\rm bol}$ is no longer contained in the equation. 
Assuming $R_{\rm in} = R_{\rm M}$ (i.e., using Equations~\ref{r_in} and \ref{r_m_edd}), 
we obtain a constraint on the disk inclination angle as: 

\begin{equation}
    \cos\theta = 5.3^{+25}_{-4.6}\qty(\frac{\kappa}{2.0})^4\qty(\frac{\xi}{0.7})^{-14/9}\qty(\frac{B_{\rm NS}}{10^{13}~\si{G}})^{-8/9}. \label{cos_edd}
\end{equation}

\begin{figure}
    \centering
    \includegraphics[scale=0.3]{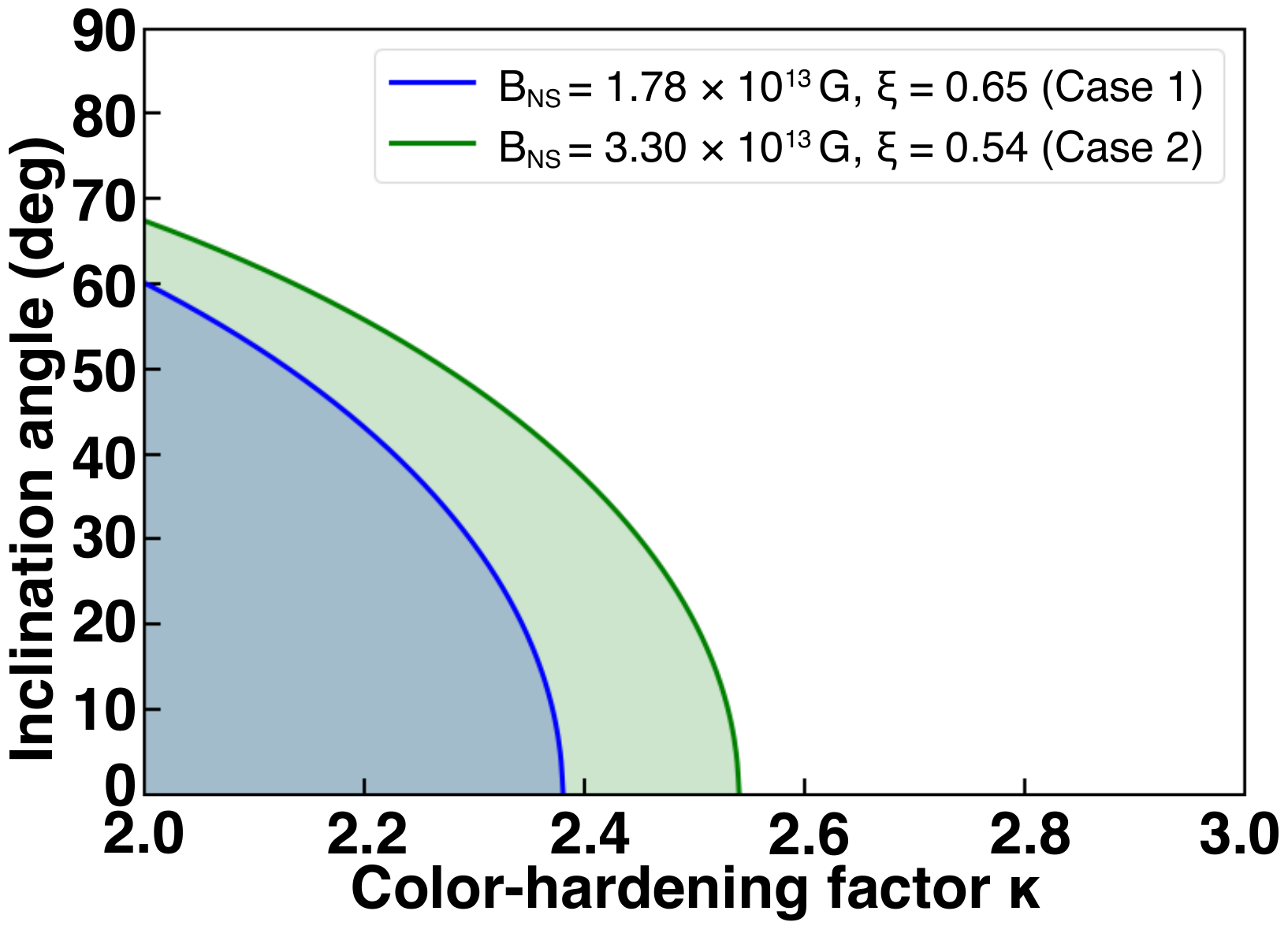}
    \caption{The inclination angle of the accretion disk as a function of the color-hardening factor $\kappa$ under different pairs of the magnetic field $B_{\rm NS}$ and $\xi$ derived in \cite{gao_magnetic_2021}. The blue and green regions indicate the allowed range of the inclination angle with a 1$\sigma$ confidence level derived from Equation \ref{cos_edd}.}
    \label{fig:kappa_vs_inclination}
\end{figure}

The model of \cite{gao_magnetic_2021} also allows us to estimate the magnetic field self-consistently. For NGC 5907 ULX1, they used the spin period and other physical quantities reported in \cite{israel_accreting_2017} (i.e., the 2014 observations that have been revisited in this work) and obtained the possible combinations of $B_{\rm NS}$ and $\xi$ as follows: 
 $(B_{\rm NS}\,{\rm [G]}, \xi) = (\SI{6.45e11},\ 0.88)$, $(\SI{6.41e11},\ 0.88)$, $(\SI{1.78e13},\ 0.65)$, and $(\SI{3.30e13},\ 0.54)$; see Table~2 of \cite{gao_magnetic_2021}. 
The first two combinations indicate  low magnetic field, yielding no solution for the inclination angle in Equation~\ref{cos_edd}. 
The other two combinations with the higher magnetic field allow a range of inclination angles as illustrated in Figure \ref{fig:kappa_vs_inclination}.
Similar to the previous discussion, a relatively low inclination and moderate $\kappa$ values are favored. 


\subsection{Accretion Geometry within the Magnetosphere}

Within the magnetosphere, the accreting matters follow the magnetic field lines toward the magnetic pole of the neutron star \citep{basko_radiative_1975,basko_limiting_1976}. In this section, we discuss the possible structure of the accretion flow that is responsible for the properties observed in the phase-variable cutoff power-law component, 
i.e., a harder spectrum observed in the pre-peak phase. 

In the case of super-critical accretion, 
the optically thick region within the magnetosphere is 
thought to become geometrically broad, presumably spreading across the regions between the neutron star's 
magnetic pole and the innermost radius of the accretion disk. 
There are two possible radiation mechanisms that are responsible for the pulsed power-law component: 
Comptonization of soft photons originating from the neutron star surface and/or the inflowing gas (e.g., \citealt{inoue_modeling_2023}) or the multi-temperature blackbody radiation from the optically thick accretion curtain \citep{mushtukov_optically_2017}.
In the former case, the photon index and cutoff energy of the power-law component can be considered to represent the average number of scatterings and electron temperature, respectively. In the latter case, on the other hand, these parameters would be related to the apparent temperature gradient and maximum temperature of the accretion flow. In either case, the parameters of the cutoff power law component can be considered to reflect the temperature structure of the accretion flow within the magnetosphere. Therefore, we can attribute the phase-variable spectral properties to an axial-asymmetric temperature distribution of the accretion flow such that a relatively hot region emerges in the pre-peak phase.

\begin{figure}
    \centering
    \includegraphics[scale=0.3]{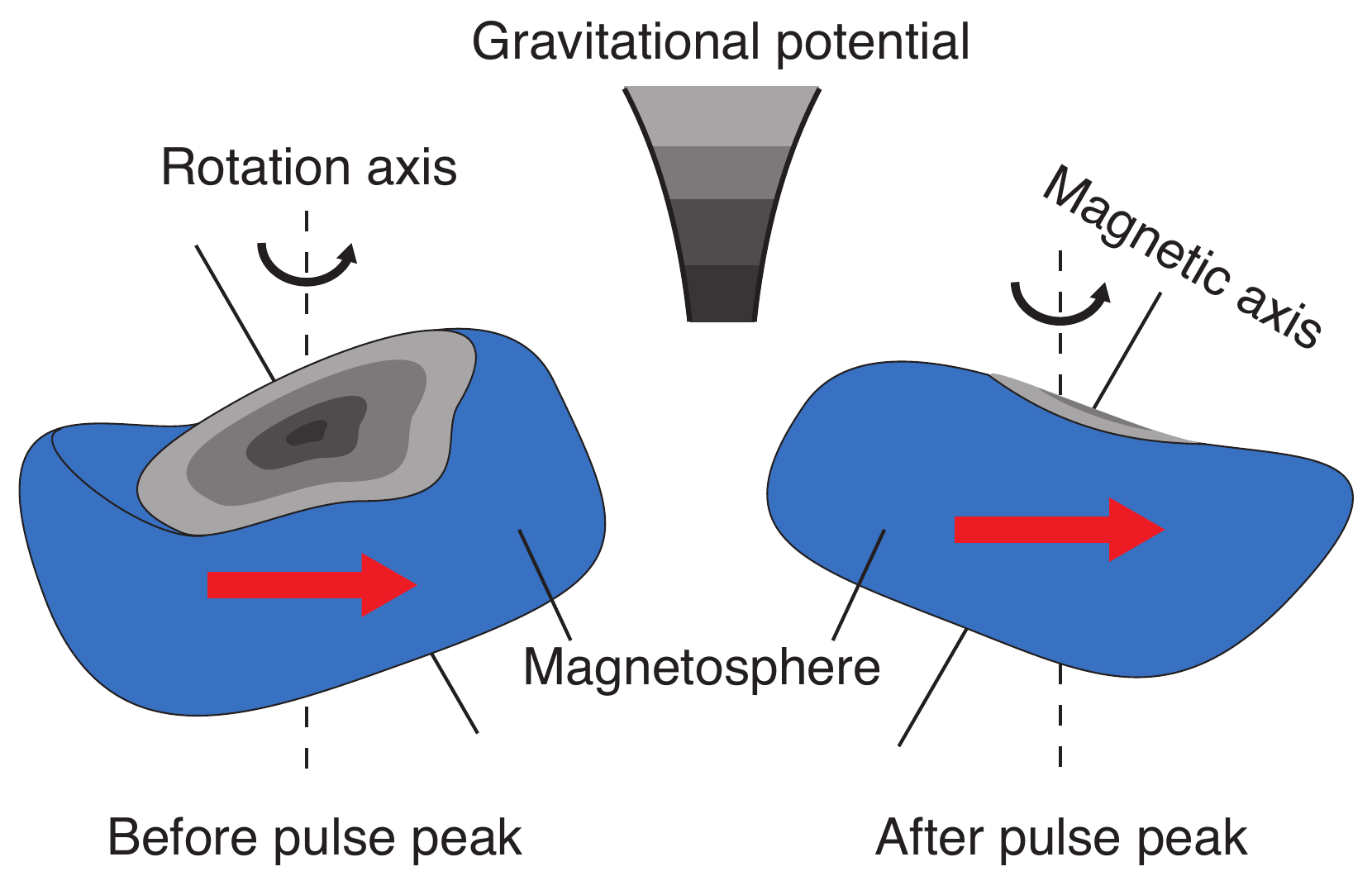}
    \caption{Schematic illustration of the possible accretion geometry within the magnetosphere. The red arrows indicate the rotational direction of the accretion disk surrounding the magnetosphere. The gray-scale gradient represents the depth of the gravitational potential.}
    \label{fig:structure}
\end{figure}

Here we propose a possible geometry of the accretion flow as illustrated in Figure~\ref{fig:structure}. 
We assume that the rotation axes of the neutron star and the accretion disk are well aligned with each other, and that the system is observed in the low inclination angle (e.g., $\theta \approx 30^\circ$). 
When the magnetic axis is oriented toward the observer, the largest cross section of the magnetosphere is achieved and thus the pulse profile peaks. 
Since the neutron star is in the spin-up phase ($\dot{P} < 0$), the magnetic field lines could be dragged in the direction of rotation by the accretion disk (red arrows in Figure \ref{fig:structure}). 
As a result, the regions closer to the magnetic pole (where the temperature is relatively high) can be more easily observed in the pre-peak phase than in the post-peak phase, making the spectrum slightly harder in the former. 
If this hypothesis is true, similar spectral behavior might be observed in other ULXPs in the spin-up phase. Therefore, a systematic study of ULXPs based on the phase-resolved spectral analysis is highly encouraged. 
Future numerical work is also urged to confirm if the proposed geometry is indeed realized and the observed spectral properties can be quantitatively reproduced. 

\section{Conclusions}\label{sec:conclusion}

In this paper, we have presented the timing and spectral analysis of the broadband X-ray data of NGC 5907 ULX1 obtained from the simultaneous observations with {\rm XMM-Newton} and {\rm NuSTAR} during its ultraluminous phase in July 2014. 
The spin period and its derivative are determined to be $P \sim 1.1$\,s and $\dot{P} \sim -5 \times 10^{-9}$\,s\,s$^{-1}$, respectively. 
The energy-dependent phase shift is observed in the folded light curves, confirming the previous results of \cite{israel_accreting_2017}. 


The emission from the accretion disk is reproduced by a multicolor disk blackbody model with an innermost disk temperature of $\sim \SI{0.3}{keV}$. 
Our measurement indicates a relatively low inclination angle of the binary system, in contrast to the previous claim by \cite{furst_spectral_2017}. 
Our result naturally explains the detection of the pulsation as well as the absence of an eclipse phase in this system. 

The spectrum of the pulsed emission component is modeled by a power law with an exponential cutoff. 
Its photon index and cutoff energy can be considered to reflect the temperature structure of the optically-thick accretion flow inside the magnetosphere.
Our phase-resolved spectral analysis indicates that this component is slightly harder in the pre-peak phase than in the post-peak phase. 
This suggests that the magnetosphere has an asymmetric geometry because of the dragged magnetic field lines, and the regions closer to the magnetic pole emerge in the pre-peak phase. A systematic observations of other ULXPs as well as numerical work are highly encouraged to confirm if this hypothesis is true. 

\section*{Acknowledgments}
The authors would like to thank the anonymous referee for her/his fruitful feedback, which helped to improve the manuscript. We would like to thank Akihiro Inoue for the useful discussion. We also thank the XMM-Newton and NuSTAR teams for their devotion to instrumental calibration and spacecraft operation.

%

\facilities{XMM, NuSTAR}


\software{NumPy \citep{harris_array_2020}, Astropy \citep{collaboration_astropy_2018}, Matplotlib \citep{4160265}, HENDRICS \citep{bachetti_maltpynt_2015}, Veusz (\url{https://veusz.github.io}), HEASoft (\url{https://heasarc.gsfc.nasa.gov/docs/software/heasoft/}), SAS (\url{https://www.cosmos.esa.int/web/xmm-newton/download-and-install-sas}), XSPEC \citep{arnaud_xspec_1996}}



\clearpage
\appendix
\section{Spectra from all eight phases and corresponding best-fit parameters}\label{sec:appendix}
\begin{figure}[h]
    \gridline{
    \fig{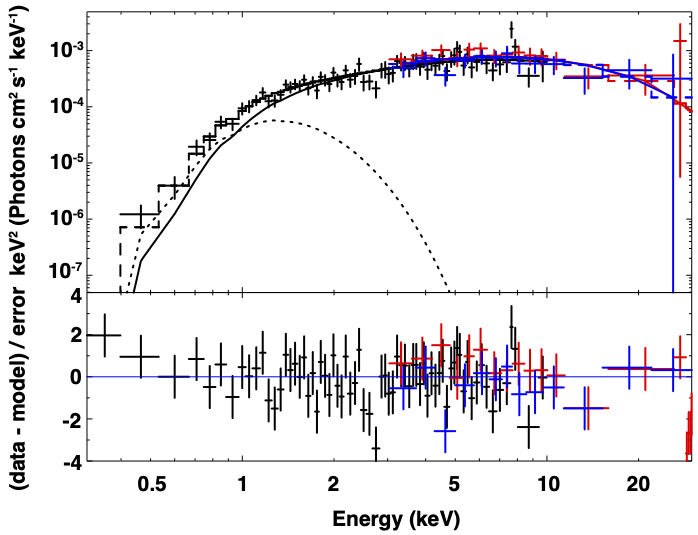}{0.33\textwidth}{(a) Phase 0.0-0.1}
    \fig{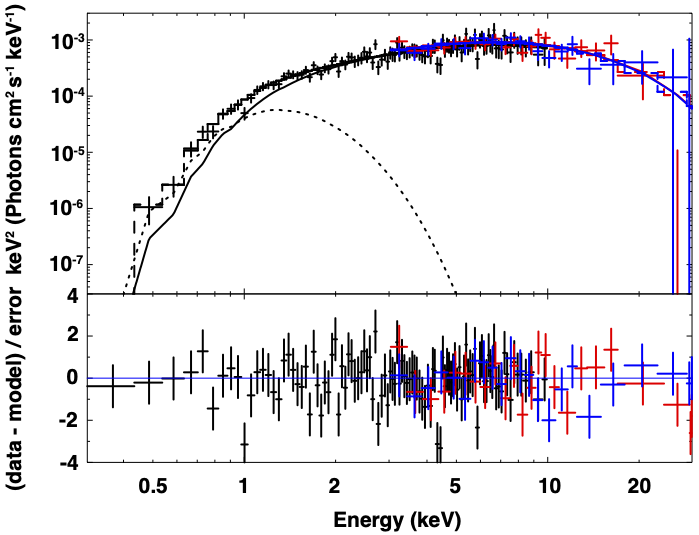}{0.33\textwidth}{(b) Phase 0.1-0.3}
    \fig{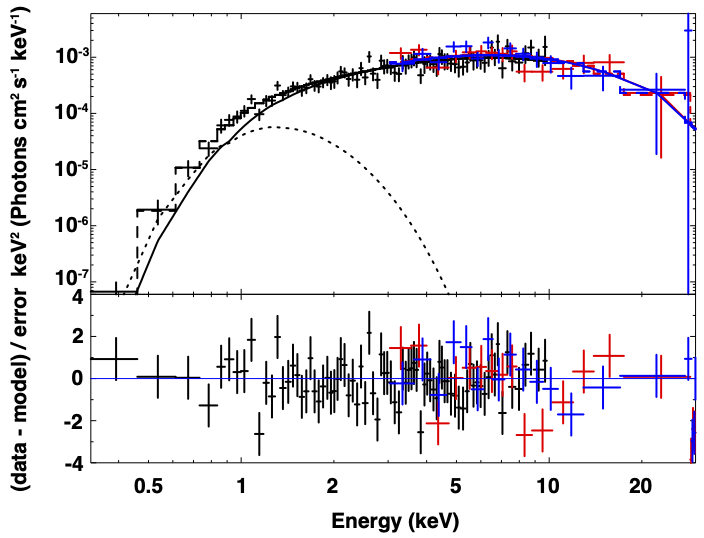}{0.33\textwidth}{(c) Phase 0.3-0.4}
    }
    \gridline{
    \fig{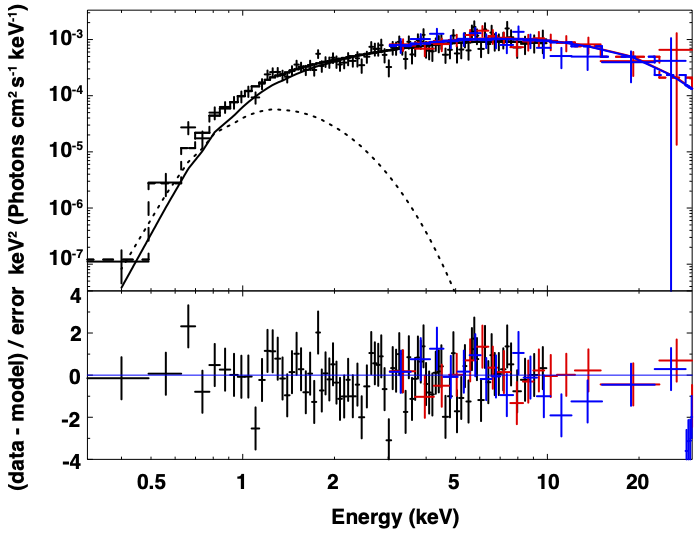}{0.33\textwidth}{(d) Phase 0.4-0.5}
    \fig{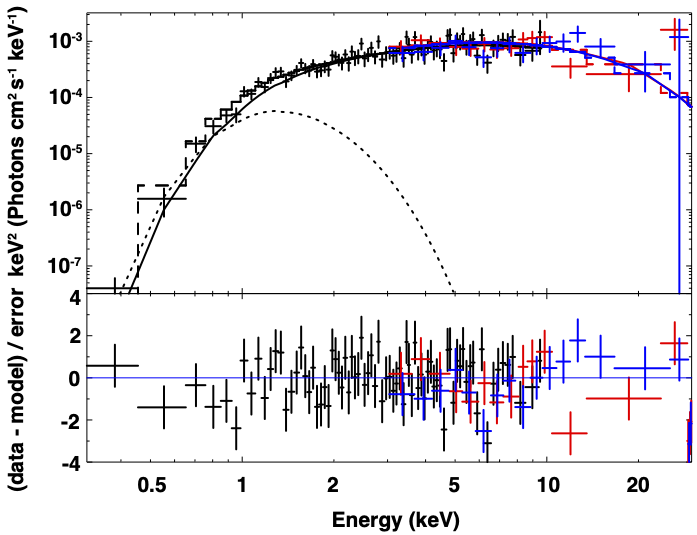}{0.33\textwidth}{(e) Phase 0.5-0.6}
    \fig{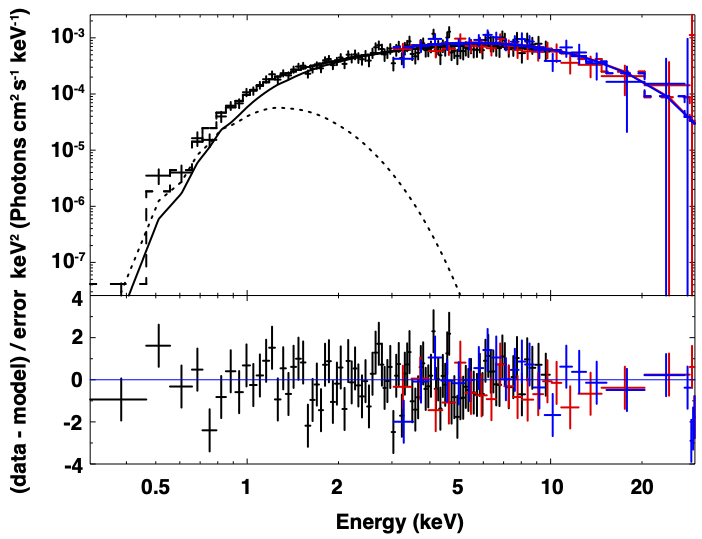}{0.33\textwidth}{(f) Phase 0.6-0.8}
    }
    \gridline{
    \fig{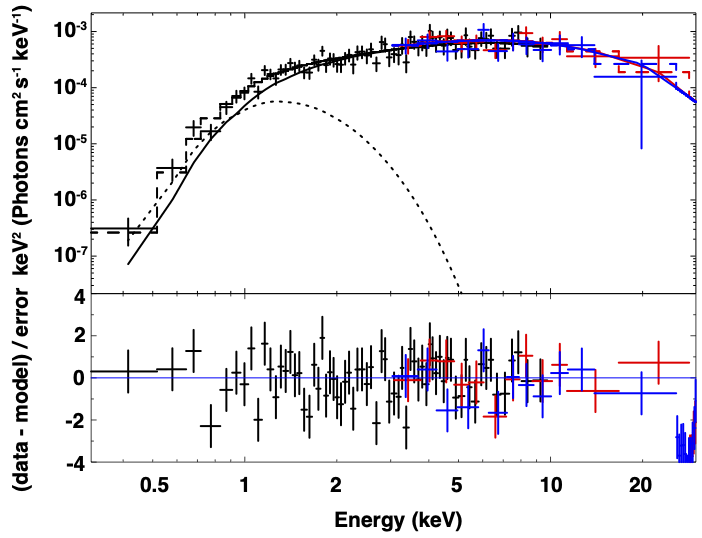}{0.33\textwidth}{(g) Phase 0.8-0.9}
    \fig{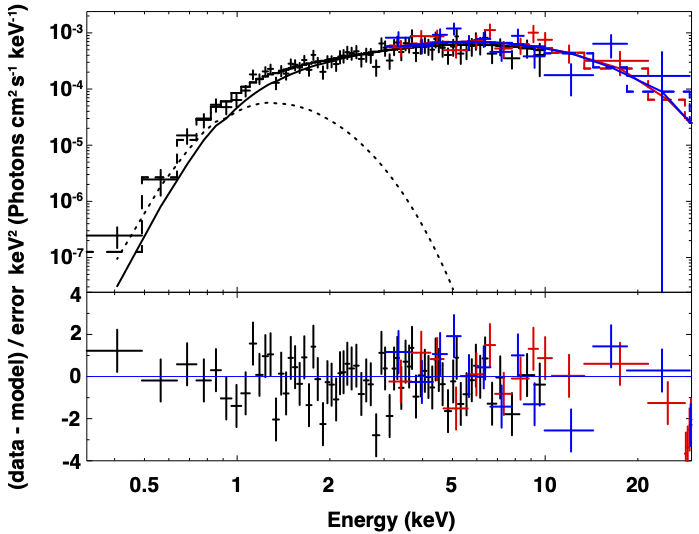}{0.33\textwidth}{(h) Phase 0.9-1.0}
    }
    \caption{Unfolded spectrum from each phase. {\rm XMM-Newton} EPIC-pn data are shown in black, {\rm NuSTAR} FPMA data are shown in red, and FPMB data are shown in blue. The best-fit \texttt{diskpbb+cutoffpl} model with $p = 0.5$ is superimposed. The dashed lines and the solid lines in each color represent the total model and the \texttt{cutoffpl} component, respectively. The black dotted lines represent the \texttt{diskpbb} component.}
    \label{fig:appendix_spectrum}
\end{figure}

Figure \ref{fig:appendix_spectrum} shows the phase-resolved spectra obtained in the manner described in Section \ref{subsec:spectral_analysis}.

In Table \ref{tab:parameter_full} we present the complete best-fit parameters. The method of the analysis from which these parameters are obtained is also described in section \ref{subsec:spectral_analysis}.

\clearpage

\begin{table}[htb!]
    \begin{center}
        \caption{Best-fit Parameters of Pulsed Emission Component for All the Spectra\label{tab:parameter_full}}
        \begin{tabular}{lccc}\hline\hline
        Phase & $\Gamma$ & $E_\mathrm{fold}~(\si{keV})$ & $F~(\SI{e-12}{erg.cm^{-2}.s^{-1}})$ \\ \hline
        $p = 0.5$ & ~ & ~ & ~ \\
        $0.0-0.1$ & $0.82^{+0.17}_{-0.19}$ & $5.8^{+1.4}_{-1.0}$ & $2.7 \pm 0.1$ \\
        $0.1-0.3$ & $0.67^{+0.18}_{-0.20}$\tablenotemark{a} & $4.9^{+0.9}_{-0.7}$\tablenotemark{a} & $3.0\pm 0.1$ \\
        $0.3-0.4$ & $0.59^{+0.15}_{-0.16}$ & $4.5^{+0.7}_{-0.6}$ & $3.5\pm 0.1$ \\
        $0.4-0.5$ & $0.90\pm0.14$ & $6.2^{+1.2}_{-0.9}$ & $3.7\pm 0.1$ \\
        $0.5-0.6$ & $0.84^{+0.15}_{-0.16}$ & $5.3^{+1.0}_{-0.8}$ & $3.3\pm 0.1$ \\ 
        $0.6-0.8$ & $0.82^{+0.18}_{-0.19}$\tablenotemark{a} & $4.6^{+0.9}_{-0.7}$\tablenotemark{a} & $2.8\pm 0.1$ \\
        $0.8-0.9$ & $0.89^{+0.16}_{-0.18}$ & $5.5^{+1.3}_{-0.9}$ & $2.5\pm 0.1$ \\
        $0.9-1.0$ & $0.77^{+0.18}_{-0.19}$ & $4.5^{+0.9}_{-0.7}$ & $2.4\pm 0.1$ \\ \hline
        $p = 0.75$ & ~ & ~ & ~ \\
        $0.0-0.1$ & $0.83^{+0.16}_{-0.18}$ & $5.8^{+1.4}_{-1.0}$ & $2.7 \pm 0.1$ \\
        $0.1-0.3$ & $0.66^{+0.19}_{-0.21}$\tablenotemark{a} & $4.9^{+1.0}_{-0.8}$\tablenotemark{a} & $3.0\pm 0.1$ \\
        $0.3-0.4$ & $0.58^{+0.15}_{-0.16}$ & $4.5^{+0.7}_{-0.6}$ & $3.5\pm 0.1$ \\
        $0.4-0.5$ & $0.88^{+0.13}_{-0.14}$ & $6.2^{+1.2}_{-0.9}$ & $3.7\pm 0.1$ \\
        $0.5-0.6$ & $0.81^{+0.15}_{-0.16}$ & $5.1^{+1.0}_{-0.7}$ & $3.3\pm 0.1$ \\ 
        $0.6-0.8$ & $0.80^{+0.20}_{-0.21}$\tablenotemark{a} & $4.5^{+1.0}_{-0.7}$\tablenotemark{a} & $2.8\pm 0.1$ \\
        $0.8-0.9$ & $0.89^{+0.16}_{-0.18}$ & $5.5^{+1.3}_{-0.9}$ & $2.5\pm 0.1$ \\
        $0.9-1.0$ & $0.77^{+0.18}_{-0.19}$ & $4.5^{+0.9}_{-0.7}$ & $2.4\pm 0.1$ \\ \hline
        \end{tabular}
    \end{center}
    \tablecomments{}
    \tablenotetext{a}{Uncertainties were obtained using XSPEC \texttt{steppar} command in the 2D-space.}
\end{table}

\clearpage




\bibliography{reference}
\bibliographystyle{aasjournal}

\end{document}